\documentclass[fleqn,usenatbib]{mnras}

\usepackage{mathptmx}

\usepackage[T1]{fontenc}

\DeclareRobustCommand{\VAN}[3]{#2}
\let\VANthebibliography\thebibliography
\def\thebibliography{\DeclareRobustCommand{\VAN}[3]{##3}\VANthebibliography}

\usepackage{graphicx}	
\usepackage{amsmath}	
\usepackage{amssymb}	

\newcommand{\new}[1]{{\color{black}#1}}
\newcommand{\msun}{M$_{\odot}$}
\usepackage{xcolor}

\title[GCs in TNG50]{Modeling globular clusters in the TNG50 simulation: predictions from dwarfs to giant galaxies}

\author[J. E. Doppel et al.]
{Jessica E. Doppel$^{1}$\thanks{E-mail: jdopp001@ucr.edu},
Laura V. Sales$^{1}$,
Dylan Nelson$^{2}$,
Annalisa Pillepich$^{3}$,
Mario G. Abadi$^{4,5}$,
Eric W. Peng$^{6,7}$,
\newauthor Federico Marinacci$^{8}$,
Jill Naiman$^{9}$,
Paul Torrey$^{10}$,
Mark Vogelsberger$^{11}$,
Rainer Weinberger$^{12}$ and
\newauthor 
Lars Hernquist$^{13}$
\\
$^{1}$University of California, Riverside, 900 University Ave, Riverside, CA 92521, USA\\
$^{2}$Institut fur theoretische Astrophysik, Zentrum fur Astronomie, Universitat at Heidelberg, D-69120 Heidelberg, Baden-Wurttemburg, Germany\\
$^{3}$Max-Planck-Institut fur Astronomie, Konigstuhl 17, 69117 Heidelberg, Germany\\
$^{4}$Instituto de Astronomia Teorica y Experimental, CONICET-UNC, Laprida 854, X5000BGR, Cordoba, Argentina\\
$^{5}$Observatorio Astronómico, Universidad Nacional de Córdoba, Laprida 854, X5000BGR, Córdoba, Argentina\\
$^{6}$Department of Astronomy, Peking University, Beijing 100871, China\\
$^{7}$ Kavli Institute for Astronomy and Astrophysics, Peking University, Beijing 100871, China\\
$^{8}$Department of Physics and Astronomy ”Augusto Righi”, University of Bologna, I-40129 Bologna, Italy\\
$^{9}$School of Information Sciences, University of Illinois, 501 E.
Daniel St, Champaign, IL 61820, USA\\
$^{10}$Department of Astronomy, University of Florida, 211 Bryant Space Science Center,Gainesville, FL 32611, USA.\\
$^{11}$Department of Physics, Massachusetts Institute of Technology, Cambridge, MA 02139, USA\\
$^{12}$Canadian Institute for Theoretical Astrophysics, 60 St. George Street, Toronto, ON M5S 3H8, Canada\\
$^{13}$Institute for Theory and Computation, Harvard-Smithsonian Center for Astrophysics, Cambridge,
MA 02138, USA.\\
}

\date{Accepted XXX. Received YYY; in original form ZZZ}

\pubyear{2022}

\begin{document}
\label{firstpage}
\pagerange{\pageref{firstpage}--\pageref{lastpage}}
\maketitle

\begin{abstract}
We present a post-processing catalog of globular clusters (GCs) for the $39$ most massive groups and clusters in the TNG50 simulation of the IlllustrisTNG project  (virial masses $M_{200} =[5\times 10^{12} \rm - 2 \times 10^{14}$] \msun). We tag GC particles to all galaxies with stellar mass $M_* \geq 5\times10^6$ \msun, and we calibrate their masses to reproduce the observed power-law relation between GC mass and halo mass for galaxies with $M_{200} \geq 10^{11}$ \msun\; (corresponding to $M_* \sim 10^9$ \msun). Here we explore whether an extrapolation of this $M_{\rm GC}$-$M_{200}$ relation to lower-mass dwarfs is consistent with current observations. We find a good agreement between our predicted number and specific frequency of GCs in dwarfs with $\rm M_*=[5 \times 10^6 \rm - 10^9]$ \msun\;  and observations. Moreover, we predict a steep decline in the GC occupation fraction for dwarfs with $M_*<10^9$ \msun\; which agrees well with current observational constraints. This declining occupation fraction is due to a combination of tidal stripping in all dwarfs plus a stochastic sampling of the GC mass function for dwarfs with $M_* < 10^{7.5}$ \msun. Our simulations also reproduce available constraints on the abundance of intra-cluster GCs in Virgo and Centaurus A. These successes provide support to the hypothesis that the $M_{\rm GC}$-$M_{200}$ relation holds, albeit with more scatter, all the way down to the regime of classical dwarf spheroidals in these environments. Our GC catalogs are publicly available as part of the IllustrisTNG data release. 
\end{abstract}




\section{Introduction}
\label{sec:intro}

The formation of globular clusters (GCs) in connection to galaxies and their dark matter halos is still unclear. Currently, the most successful models link the formation of GCs (or their early progenitors) to baryonic processes in the interstellar medium (ISM) of  galaxies. These processes are connected to star formation in high density/high pressure environments \citep{kruijssen2012, kruijssen2015, elmegreen2017} and best sampled in mergers and early stages of galaxy formation \citep{kravtsov2005,Prieto2008,Li2014,Renaud2015}. However, GCs have also been hypothesized to form at the centers of their own low-mass dark matter halos before reionization \citep{Peebles1984, Boylan-Kolchin2017}, later infalling onto larger galaxies and groups and cluster halos to form the clustered GC distributions typically found in these systems \citep{Diemand2005,Creasey2019}. Although this scenario predicts older ages and lower metallicities for GCs than current measurements \citep{lotz2004,Bastian2020}, the discovery of a few very metal poor GCs in M31 and the Milky Way may provide some support to such pristine formation scenarios playing at least some role in building the population of GCs observed in galaxies today \citep{Larsen2020,Martin2022,Errani2022}.

Observationally, the mass in GCs is found to be a power-law function of inferred halo mass for galaxies with stellar mass $M_* \geq 10^{10}$ \msun\; \citep{blakeslee1997, peng2008, spitler2009, georgiev2010, harris2013, hudson2014a, harris2015}. At face value, this relation may encode important information on the formation scenario of GCs. Theoretical models suggest that a quasi-linear power-law relation between GC mass and halo mass may arise naturally in hierarchical formation scenarios as the result of consecutive mergers, serving more as a confirmation of the hierarchical assembly of galaxies rather than shedding light on the formation mechanism of GCs themselves \citep{El-Badry2019}. However, in the regime of dwarf galaxies ($M_* \leq 10^9$ \msun), there are fewer mergers with GC-bearing companions, offering a clearer window for the study of GC formation mechanisms than in more massive galaxies. It is therefore important to extend the study of the GC mass - halo mass relation to lower-mass galaxies.

Theoretical models linking the formation of GCs to the ISM of galaxies seem to suggest a downturn in the efficiency of GC formation in dwarfs, departing from the extrapolation of the GC mass - halo mass relation measured on more massive galaxies \citep{El-Badry2019, Choksi2019,Bastian2020}. The lower efficiency of GC formation per halo mass in dwarf galaxies is naturally expected due to the lower baryonic content in low-mass halos, which limits the available gas to form stellar clusters in merger and accretion events. On the other hand, a scenario where GCs are linked to dark matter mini-halos would imply a single power-law relation between GC mass and halo mass in the regime of dwarfs, due to the self-similarity of subhalo mass in $\Lambda$CDM \citep[e.g., ][]{Creasey2019}. Although current observational constraints on the radial distribution combined with the abundance of GCs in MW-mass galaxies  limits the fraction of GCs formed in mini-halos to $\sim 30\%$ \citep{Creasey2019} for such hosts, the importance of the mini-halo formation scenario for GCs in the regime of dwarfs remains largely unconstrained.

Measuring the relation between GC mass and halo mass on the scale of dwarfs is, however, very challenging. First, while there are several methods to estimate halo mass from observables in more massive galaxies (lensing, rotation curves, abundance matching), halo mass estimates in the scale of dwarf galaxies are more scarce and uncertain. Second, GC numbers are lower in low-mass galaxies, meaning that completeness and contamination in GC surveys impact more heavily low-mass dwarfs than estimates for high-mass galaxies. There are, however, several observational efforts to constrain the GC content in dwarf galaxies. Most notably, \citet{forbes2018} finds that dwarfs in the Local Volume are consistent with an extrapolation of the power law relation between GC mass and halo mass observed in more massive galaxies, where halo masses for the dwarf galaxies are estimated using gas kinematics. However, other work cautions that this might be biased to include only dwarfs that have at least one GC, while including all dwarfs of a given mass in the average could lead to a departure downwards from the power-law extrapolation \citep{Bastian2020}. 

In light of this discussion, another important diagnostic emerges as a potential constraint: the ability of galaxies of a given mass to host at least one GC, or the GC occupation fraction. Observations in the Virgo cluster suggest that all dwarfs with $M_{*} > 10^9$ M$_{\odot}$ have GCs, but that fraction declines quite steeply for lower-mass objects, finding $50\%$ occupation in dwarfs with $M_{*} \sim 10^{7.5}$  M$_{\odot}$ \citep{sanchezjanessen2019}, which is similar to the conclusion presented in \citet{Eadie2022} using a compilation of available data for dwarfs. Recently a comparable occupation fraction was reported for dwarf galaxy satellites of MW-like primaries in the local volume \citep{Carlsten2022}. However, the available constraints involve mostly satellite dwarfs, or dwarf galaxies embedded in the gravitational potential of larger hosts, meaning that tidal stripping and other environmental effects might have influenced their original GC content, preventing a simple interpretation. Unfortunately, surveys of GCs in field dwarfs are scarce and still insufficient to constrain GC occupation fractions (e.g, \citealt{georgiev2010}).

An interesting path forward is to use cosmological simulations of dwarf galaxies in high density environments to understand the connection between GCs, dwarf galaxies and their dark matter halos.  This is particularly appealing since  hydrodynamical cosmological simulations of representative volumes of the Universe have been powerful tools to understand and model the evolution of satellite dwarfs and their properties -- such as color, mass content, morphology -- in the environments of groups and clusters \citep{Sales2015, Yun2019, Joshi2020, Vogelsberger2020, Donnari2021a, Engler2021a, Joshi2021} creating a realistic population of satellite dwarfs in good agreement with observations \citep{Donnari2021b, Engler2021b, Riggs2022}. However, the spatial and mass resolution of such simulations is too coarse to directly resolve the process of GC formation.

While employing idealized galaxy and galaxy merger set-ups \citep{Bekki2002,Kruijssen2012b,Renaud2015,lahen2019, Lahen2020, li2021} or cosmological zoom-in of galaxies at high redshifts \citep{kim2017, Ma2020, Sameie2022} have shown important successes on simulating the formation and evolution of GCs and their connection to the ISM of the host galaxy, these techniques are currently unable to sample the evolutionary history of galaxies until the present day and within high density environments, where most of the GC observational data is available today.

To circumvent this limitation, in this paper we develop a GC catalog added in post-processing via a particle tagging technique to make predictions on the abundance, distribution and kinematics of surviving GCs in the environments of groups and clusters at $z = 0$. This technique is inspired by the successes of particle-tagging for studying stellar halo science \citep{Bullock2005,Penarrubia2008, Cooper2010, Laporte2013} and it has been shown to have success in modeling the {\it surviving} population of GCs in cosmological simulations of galaxy clusters \citep{ramos2015, mistani2016, ramosalmendares2018, ramosalmendares2020, doppel2021}. 

Tagging techniques of this kind mentioned above are complementary to more detailed methods where GCs formation sites are identified in hydrodynamical simulations and followed in time by a set of sub-grid prescriptions to model their evolution until the present day \citep[e.g., ][]{kruijssen2011, mistani2016, Li2017,pfeffer2018, Keller2020, Trujillo-Gomez2021, Chen2022, Reina-Campos2022}. Note that most of these works also require fairly high-resolution simulations and have been mostly focused on the scale of MW-like galaxies so far. Instead, the less computationally-intensive modeling associated with particle tagging methods offer the opportunity to compile theoretical predictions for the GC content, their positions and velocities for a large number of galaxies and dwarfs with realistic properties within high density environments such as simulated groups and galaxy clusters.

Here we extend the particle-tagging method applied in \citet{ramosalmendares2020} for Fornax and Virgo mass galaxy clusters ($M_{200} \geq 8 \times 10^{13}$ M$_{\odot}$) that was implemented in the Illustris simulations \citep{vogelsberger2013, vogelsberger2014a, vogelsberger2014b, genel2014} to lower-mass dwarf galaxies using the highest-resolution hydrodynamical run of the TNG50 simulation \citep{pillepich2019, nelson2019}.  Thus, the tagged GCs allow us to study the GC content of a variety of galaxy groups and clusters consistent with mass estimates of Centaurus A, Fornax, Hydra and Virgo, where observations of GCs are abundant. This work presents one of the largest studies of its kind, containing $39$ groups and clusters including their associated $5000+$ galaxies with $M_* \geq 5\times 10^6$ \msun, and $196,611$ GCs. The GC catalogs generated for this work are made publicly available\footnote{www.tng-project.org/doppel22} (see Data Availability section for accessibility information).

The paper is organized as follows.  In Sec.~\ref{sec:methods}, we present the simulation and GC tagging technique. In Sec.~\ref{sec:icgc} we show our results on the  intra-cluster GC component and benchmark our catalog using current observations. Our main results on the content of GCs in dwarf galaxies are shown in Sec.~\ref{sec:GCs_in_galaxies} and Sec. \ref{sec:occupation_fraction}. We summarize our main findings in Sec.~\ref{sec:summary}.

\section{Methods}
\label{sec:methods}
\subsection{The TNG50 Simulation}
\label{ssec:sims}

We use the highest-resolution hydrodynamical run of the TNG50 simulation \citep{pillepich2019, nelson2019}, which allows us to relate the properties of the tagged GCs directly to the properties of galaxies, galaxy groups or galaxy clusters that they belong to. TNG50 is an unprecedentedly high resolution cosmological hydrodynamical simulation for its volume, with a box size of $51.7$ Mpc per side with $2160^3$ gas and dark matter particles, allowing for a mass resolution of, on average, $8.4 \times 10^4$ M$_{\odot}$ for baryons and a fixed mass resolution of $4.5\times10^5$ M$_{\odot}$ for dark matter. The simulation has a gravitational softening length of $288$ pc for stars and dark matter at $z = 0$. TNG50 assumes a flat, $\Lambda$CDM cosmology and uses cosmological parameters from \citet{planckcollab2016}.  Its galaxy formation model follows star formation in moderately dense ISM conditions, stellar evolution and chemical enrichment via supernovae, primordial and metal line cooling of gas, as well as heating from the background radiation field, the seeding and subsequent growth of supermassive black holes as well as AGN feedback at both low and high accretion rates, and galactic winds \citep{weinberger2017, pillepich2018}. The TNG50 simulation is part of the larger IllustrisTNG project \citep{naiman2018, pillepich2018b, nelson2018, springel2018, marinacci2018, nelson2019dr}. 


\subsection{Galaxy Selection}
\label{ssec:selection} 

\begin{figure}
    \centering
    \includegraphics[width = \columnwidth]{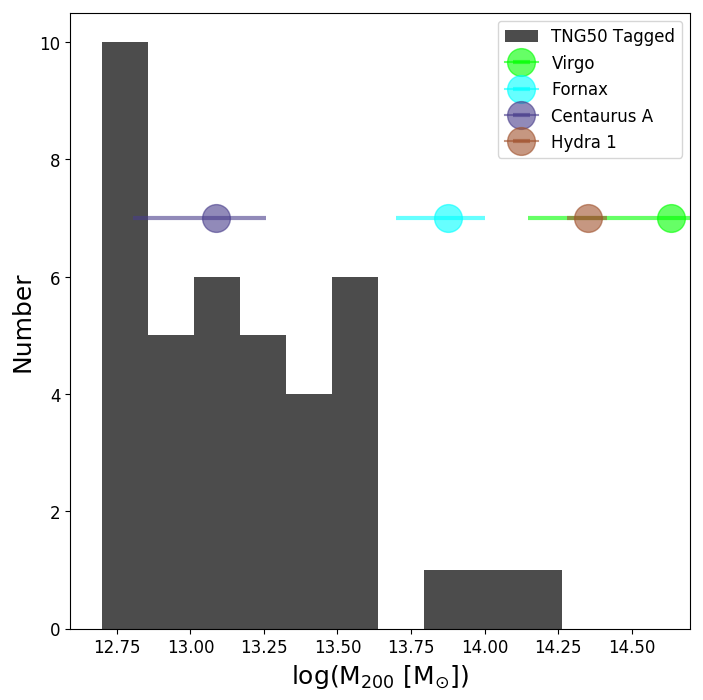}
    \caption{The distribution of TNG50 $z = 0 $ virial masses ($M_{200}$) of the 39 most massive galaxy groups and clusters within the simulation to which we tag GCs (black histogram). We cover a wide range of masses, from Centaurus A on the low-mass end \citep[dark purple circle and errorbar ][]{vandenbergh2000, karachentsev2007}, to Fornax \citep[cyan circle and errorbar ][]{drinkwater2001}, and to Hydra 1 \citep[brown circle and errorbar ][]{tamura2000} and low end mass estimates of Virgo \citep[lime green circle and errorbar ][]{karachentsev2010, Weinmann2011} on the high-mass end. The $\sim 1.5$ dex range of virial masses allows us to study potential effects environment might play in their $z = 0$ GCs.}
    \label{fig:mass_hist}
\end{figure}

We tag GCs in all TNG50 host halos with a virial mass M$_{200} \geq 5\times 10^{12}$ M$_{\odot}$, (where $M_{200}$, refers to the mass within the virial radius $r_{200}$ defined as the radius enclosing an average density equal to $200$ times the critical density of the Universe). This selection results in $39$ groups and clusters with a virial mass distribution shown in Fig.~\ref{fig:mass_hist}. The high-mass end is roughly on par with lower estimated virial masses of the Virgo cluster $M_{200} \sim 10^{14} $ M$_{\odot}$  \citep[lime green circle and errorbar ][]{karachentsev2010, Weinmann2011} and Hydra 1 \citep[brown circle and errorbar][]{tamura2000}, Fornax cluster $M_{200} \sim 10^{14}$ M$_{\odot}$ \citep[cyan circle and errorbar ][]{drinkwater2001}, down to Centaurus A with estimated $M_{200} \leq 10^{13}$ M$_{\odot}$ \citep[dark purple circle and errorbar ][]{vandenbergh2000, karachentsev2007} and less massive elliptical systems closer to the lower-mass cut. 

 
We identify all galaxies that interacted with each of these groups (defined here as being part of their merger tree) and achieved a maximum stellar mass M$_{*, \rm max} \geq 5\times 10^6$ M$_{\odot}$ during their lifetime as candidates to host the tagged GCs. For each of our selected galaxies, we calculate their infall time $t_{\rm inf}$, defined here by following their main branch progenitors in the Sublink merger tree \citep{rodriguez-gomez2015} to the last time that the progenitor was its own central. This corresponds to the snapshot before they begin interacting with their current host galaxy group or cluster, or any lower-mass halo which eventually merges with the group or cluster \citep[i.e., in pre-processing][]{Benavides2020,Joshi2021}. Here, we also impose a minimum of $100$ dark matter particles, to remove spurious objects in the subhalo catalog. In the case of the central galaxy in each of our $39$ groups, following \citet{ramosalmendares2020}, we define the infall time as the snapshot when the main progenitor branch reaches $5\%$ its $z = 0$ value.

The target selection process gives us $8746$ progenitor galaxies to be tagged with GCs at their infall time of which $6415$ survive to $z = 0$. For our study of GCs associated with satellites galaxies, our final sample includes $5453$ satellite galaxies in groups and clusters with M$_{*} \geq 5\times 10^6$ M$_{\odot}$ at $z = 0$, which guarantee well resolved galaxies with at least $\sim 60$ stellar particles at $z = 0$. 

\subsection{Globular Cluster Tagging}
\label{ssec:tag}

The method to tag GCs in our cosmological simulation follows mostly from the one already introduced in the Illustris simulations by \citet{ramosalmendares2020}, with some modifications and improvements to extend the model to lower-mass galaxies. The method ``tags" GCs to a set of dark matter particles, selected to have a given energy distribution (enforced through a specific distribution function) that matches observational properties of GCs systems at $z = 0$. In principle, one could choose to tag on any collisionless-type particle, for example stars. We instead favor the tagging of dark matter particles to ensure that all galaxies have enough available particles with the desired distribution function to select from when assigning GCs. In particular, GCs systems observed in galaxies are dispersion dominated and typically more extended than the stellar component. Using the dark matter component to search for suitable tracers in energy-space ensures that we maximize the number of candidate particles to host GCs since dark matter is always dispersion dominated (unlike stars in disks) and more extended than the stars. This is particularly important in the regime of dwarfs, where the stellar content is low resulting on a low number of stellar particles overall and even lower beyond the inner central regions \citep[see for instance declining stellar halo fractions predicted in dwarf galaxies, Fig. 5 ][]{Elias2018}. Following \citet{ramosalmendares2020}, the tagging is done only once (at infall time) for each galaxy, after which the particle ID is used to identify those tagged GCs in the $z=0$ snapshot. 

The first step is to identify, for each object, the maximum subset of dark matter particles that are candidates to be GCs, defined as those that are consistent with a specified distribution in energy adopted for the GCs. \new{We assume that the dark matter follows a NFW profile \citep{navarro1996}:
\begin{equation}
    \rho_{\rm NFW}(r) = \frac{\rho_{\rm NFW}^0}{(r/r_{\rm NFW})(1 + r/r_{\rm NFW})^2}
\end{equation}
which we find by best-fit to the density distribution of dark matter particles following \citet{lokas2001} at infall time. We assume $r_{NFW} = r_{\rm max}/\alpha$, where $r_{\rm max}$ is the radius of maximum circular velocity and $\alpha = 2.1623$ \citep{bullock2001}. We calculate $r_{\rm max}$ for each galaxy at their time of infall.} 

\new{GCs are assumed to follow a Hernquist profile \citep{hernquist1990}:
\begin{equation}
    \rho_{\rm HQ}(r) = \frac{\rho_{\rm HQ}^0}{(r/r_{\rm HQ})(1 + r/r_{\rm HQ})^3}
\end{equation}
Two sets of GCs are tagged, one corresponding to a more extended metal poor or ``blue" component, and one more concentrated and metal rich, or ``red" component, with relative fraction of red to blue component following observations in \citet{harris2015}. We assume that $r_{\rm HQ} = \beta r_{\rm NFW}$, where $\beta_{\rm blue \ GCs} = 3.0$ and $\beta_{\rm red \ GCs} = 0.5$. The remaining parameter $\rho_{\rm HQ}^0$ is fit such that the number of resultant candidate particles is maximized. The procedure as well as the assumed parameters is the same as introduced in \citet{ramosalmendares2020} using the Illustris simulations}. 

\new{For reference, the resulting radial distributions of red and blue GCs are typically within the tidal radius of surviving satellites, which are estimated to be $r_{\rm tidal} \sim 5$ - $100$ kpc in our sample using analytical calculations for our highest and lowest host and satellite masses \citep{binney2008,Springel2008}. This is confirmed by a very high fraction of tagged GCs remaining bound at $z=0$, which show medians $96\%$ and $85\%$ for red and blue GCs, respectively. Note that although the sample as a whole shows large bound fractions, a minority of individual objects may lose most or in some cases {\it all} of their GCs for specific orbits or accretion histories, introducing scatter in some of the relations explored in Sec.~\ref{sec:GCs_in_galaxies}.}

We \new{numerically} compute the distribution function of each of these three components (dark matter NFW, blue GCs and red GCs) as \citet{bandt2008}:

\begin{equation}
    f_{i}(\epsilon) = \frac{1}{8\pi}\bigg[ \int_{0}^{\epsilon} \frac{\rm d^2\rho_i}{\rm d\psi^2} \frac{\rm d \psi}{\sqrt{\epsilon - \psi}} + \frac{1}{\sqrt{\epsilon}} \bigg(\frac{\rm d\rho_i}{\rm d\psi}\bigg) \bigg|_{\psi = 0}\bigg] ,
    \label{eqn:distfunc}
\end{equation}

\noindent
where $\rho_i$ is the density profile of i = DM, GCs, blue GCs, $\Psi$ is the relative gravitational potential, and $\epsilon$ is the relative energy. Since the potential is not recorded in every snapshot, it should be noted that the potential of the dark matter particles is calculated for each progenitor subhalo via a tree gravity for computational efficiency. Then, in equally spaced bins of relative energy, we select a fraction of the particles $f_{\mathrm{HQ}, i}/f_{\rm NFW}$ for i = red and blue GCs to be the GC candidate particles. We impose and additional radius cut of $r_h / 3$, as suggested by \citet{yahagi2005} and implemented in \cite{ramosalmendares2020}, where $r_h$ is the half-mass radius of the entire halo at its infall time. This is the final set of GC \textit{candidate} particles.

The next step is to populate galaxies with a total mass in GCs, or $M_{\rm GC}$. This is the key assumption of the method: galaxies follow a power-law relation between the mass of their total GC systems and their virial mass $M_{200}$ at infall. We thus calibrate the model such that after evolving in the cluster of host potential (tidal stripping, stellar evolution), they reproduce the observed 
power-law $M_{\rm GC} - M_{200}$ relation at $z = 0$.
More specifically, from \citet{harris2015}:
\begin{equation}
    M_{\mathrm{GC}, z = 0} = a M_{\rm halo, z = 0}^b ,
     \label{eq:calibration1}
\end{equation}

\noindent
where a = $2.6 \times 10^{-8}$ and $4.9 \times 10^{-5}$ for red and blue GCs, respectively, and the slopes $b = 1.2$ and $0.96$ for red and blue GCs. As done in \citet{harris2015}, $M_{\rm halo, z = 0}$ is calculated using abundance matching parameters from \citet{hudson2015} to assign halo masses to satellites. To calibrate this relation, we select from our satellite sample described in Sec.~\ref{ssec:selection},  only those that survived to $z=0$, and calculate the fraction of the candidates GC particles that are still bound to the galaxy at $z=0$: $f_{\rm bound} = N_{\rm candidates(z=0)}/N_{\rm  candidates(z_{\rm inf})}$. We consider a GC candidate still bound to a subhalo at present day if its corresponding dark matter particle is considered bound to the subhalo via Subfind. We then make the assumption that the relationship between $M_{\rm GC} - M_{\rm halo}$ also followed a power law at infall such that:

\begin{equation}
    M_{\rm GC, inf} = \frac{1}{f_{\rm bound}} M_{\mathrm{GC}, z = 0} = a_{\rm inf} M_{\rm halo, inf}^{b_{\rm inf}} .
    \label{eq:calibration2}
\end{equation}

\begin{figure}
    \centering
    \includegraphics[width = \columnwidth]{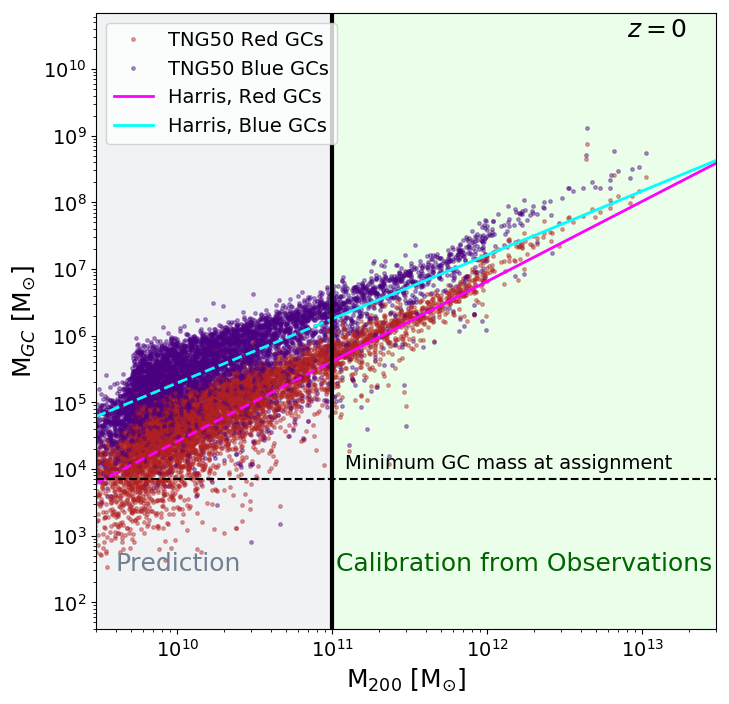}
    \caption{M$_{\mathrm{GC}}$ as a function of halo mass M$_{200}$ at present-day from the TNG50 simulation + GC tagging model which shows the result of the mass calibration process. Individual dots show simulated galaxies for the blue (indigo) and red (red) components. The observed GC mass - halo mass relation from \citet{harris2015} is shown in cyan and magenta solid lines for blue and red GCs, respectively. The extrapolation of those results to dwarf galaxies is indicated with the same colors but using short dashed lines. For galaxies with $M_{200} > 10^{12}$ M$_{\odot}$, we plot the virial mass corresponding to the simulation value rather than calculated from \citet{hudson2015} due to large discrepancies between simulations and the abundance matching model in that regime. The calibrated red and blue GCs follow a power-law with a slope in good agreement with observations and predict a variable scatter that increases towards the low-mass end. Most importantly, results for $M_{200}<10^{11}$ \msun\; are a prediction of the model since the calibration is done only using systems more massive than this cutoff. The horizontal dashed line shows $M_{\rm GC}=7\times 10^3$ \msun, our minimum individual GC mass considered to assign mass to the tagged GC particles. Galaxies below this mass are not populated with GCs in our model.} 
    \label{fig:masscal}
\end{figure}

\noindent
We find the best fitting $a_{\mathrm{inf}}$ = $2.6 \times 10^{-7}$ and $7.3 \times 10^{-5}$ and $b_{\mathrm{inf}}$ = 1.14 and 0.98 for red and blue GCs respectively. The infall GC mass of each galaxy is then calculated using their virial mass from this best-fit infall relation at $t_{\rm inf}$.
The result of this calibration is shown in Fig \ref{fig:masscal}. Blue and red points represent the resulting present-day blue and red GC mass respectively for each galaxy with a given $M_{200}$. For reference, the magenta and cyan lines show the results from \citet{harris2015} for  red and blue GCs, respectively. Note that, despite all galaxies starting from a scatter-free infall $M_{\rm GC}$-$M_{200}$ relation, the variations in infall time, tidal stripping and stellar evolution of the galaxies (which might influence the calculation of $M_{200}$ from abundance matching) results in a present-day $M_{\rm GC}$-$M_{200}$ relation with scatter, in agreement with observations \citep[see ][ for more detailed discussions]{ramosalmendares2018,ramosalmendares2020}.

Most importantly, the calibration to determine $M_{\rm GC}$ is done using only more massive galaxies, where observational constraints on the GC-halo mass relation are available. In particular, only satellites with estimated $M_{200} \geq 10^{11}$ \msun, which roughly corresponds to $M_* \geq 10^9$ \msun\; using \citet{hudson2015} are used to calibrate the model. For dwarf galaxies with  $M_{200} < 10^{11}$ \msun, the calculated $M_{\rm GC}$ is a prediction of the model assuming they follow the same relation as their more massive counterparts.

\subsection{Assigning individual GC Masses}
\label{ssec:gcmass}

As explained above, the tagging method first selects as many GC candidate particles as possible, by identifying all dark matter particles with matching energies to the intended GC distribution (see Sec.~\ref{ssec:tag}). After the mass calibration is carried out and $M_{\rm GC}$ is defined at infall (see Eqs.~\ref{eq:calibration1} and ~\ref{eq:calibration2}), the mass weight of each candidate GC particle is simply calculated by dividing $M_{\rm GC}$ into the identified number of candidate GC particles. This means that the weight of a given tagged GC particle could be smaller than the mass of a full GC. While working with the full set of candidate GC particles provides the most complete representation of the possible phase space for GC systems, it is convenient to define a ``realistic GC catalog", where a subset of the tagged GC candidate particles are selected to match the {\it number} of GCs expected. We take this approach in what follows as it allows a consistent comparison to observational data.

In previous iterations of this GC tagging model, we have taken the approach of assigning all realistic GCs the same, average mass ($m_{\rm GC} = 1\times 10^5 $ M$_{\odot}$) \citep[see ][]{ramosalmendares2020, doppel2021}. While this approach was correct for the more massive galaxy sample presented in these previous works, the GC luminosity function in observed early-type galaxies changes with the stellar mass of the host  \citep{jordan2007}, an effect that becomes particularly important when evaluating the GC content of lower-mass dwarf galaxies \citep{forbes2018}. Because of our increased resolution and the selection of dwarf galaxies below $M_* \sim 10^8$ \msun, we enter the regime in which a more detailed mass modeling for the GCs is required. 

We model the GC population of each galaxy at infall assuming a Gaussian distribution in luminosity (we assume a mass to light ratio $(M/L)$ (M$_{\odot}$ / L$_{\odot}) = 1$ in the z-band), with a dispersion of z-band GC luminosities ($\sigma_z$) that reproduces the relationship with the $M_{B}$ of their host galaxies at $z = 0$ as measured in \cite{jordan2007} (see Appendix~\ref{app:colors} for a more detailed discussion of this calibration). We note that we still assume a constant mean luminosity $\sim 2\times 10^{5} \rm L_{\odot}$ for all GC luminosity functions, independent of the mass of the host galaxy, but we limit the maximum mass that a GC can sample to $1/100$ the stellar mass of the host following observations of the most massive GCs in dwarfs \citep{kruijssen2012}. We also employ a uniform low-mass (or luminosity) cutoff for individual GCs $=7 \times 10^{3}$ M$_{\odot}$ and an upper mass/luminosity cut off equal to $5 \times 10^6$ \msun\; to ensure that we are excluding massive objects that could be nuclear GCs \citep{kruijssen2012}.

\begin{figure}
    \includegraphics[width = \columnwidth]{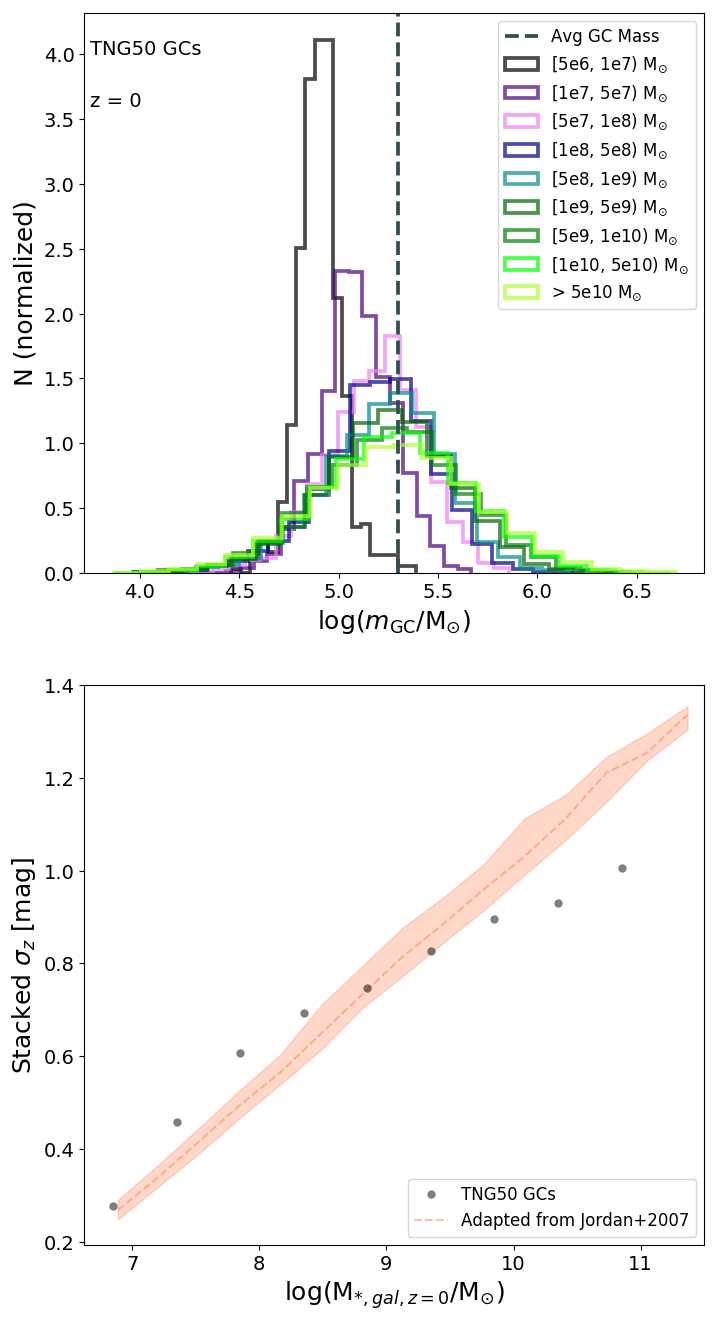}
    \caption{\textit{Top:} Stacked mass functions of individual GCs in narrow bins of host galaxy stellar mass at $z = 0$, as labelled. 
    We can quantitatively see the expected decrease in dispersion for decreasing stellar mass. We note that the downward shift of the median GC mass is due to the upper limit of $\mathrm{min}(5\times 10^6$ M$_{\odot}, M_{*, \ \mathrm{infall}}/100)$, which plays a large role for lower-mass galaxies. \textit{Bottom:} Relation between the dispersion in the z-band GC luminosity function, $\sigma_z$, and host galaxy stellar mass. Gray points show the measured dispersion of the stacked luminosity functions of our model in TNG50 shown in the top panel, and the orange shaded region shows observational expectations based on results reported in \citet{jordan2007} adapted using the simulations to convert their $B$- magnitudes to $M_*$ (median shown in dashed line, and shading corresponds to  25\% - 75\% scatter in each stellar mass bin). Additionally, due to the high end mass cut of $5\times 10^6$ M$_{\odot}$ for a single GC mass, the best-fitting luminosity function dispersions of high-mass systems are somewhat underestimated, but we do find reasonable agreement for low-mass systems.}
    \label{fig:mass_func}
\end{figure}

For each galaxy, we proceed to randomly draw individual GC masses from the resulting Gaussian distribution until the sum of all realistic GC candidates adds up to the estimated mass in GCs at infall. This steps concludes with a corresponding number of realistic GCs for each galaxy, $N_{\rm GC,inf}$. This number is always smaller than the number of particles identified as candidate GC particles in the step described in Sec.~\ref{ssec:tag}. We then sub-sample $N_{\rm GC,inf}$ from the list of all the GC candidate particles identified for each galaxy (enforcing that they follow the same relative energy distribution function) and we assign them one of the drawn GC masses, building one possible realization of the realistic GC catalog for each given galaxy. Note that further versions of the realistic catalog might be constructed by repeating the sampling of the GC luminosity function and the selection of the GC candidate particles, if so desired. In this work we employ only one realization per galaxy, but see \citet{doppel2021} for an example of how multiple realizations per object might be used to assess the impact of low number statistics in the determination of galaxy velocity dispersion from GC tracers.

As highlighted before, the individual mass assignment to GCs is performed at infall, and particle IDs are tracked onwards to $z=0$. Since the tagging technique is meant to model the \textit{surviving} GCs at $z=0$, we do not make assumptions about the shape of the {\it initial mass function} of GCs, nor do we take into account mass loss for individual GCs or the total destruction of GCs \citep[see also ][ for a detailed discussion]{ramosalmendares2020}. We instead use observational results on the evolved GC luminosity function presented in \citet{jordan2007} to assign final masses to the tagged surviving GCs. We present in Appendix \ref{app:dynamical_friction} estimates of the dynamical friction expected for the tagged GCs and demonstrate that the results presented here are not strongly affected by dynamical friction. 

\subsection{The GC population of TNG50 group and cluster members}

The top panel of Fig.~\ref{fig:mass_func} shows the average GC mass functions at $z=0$ binned in ranges of stellar mass of the host galaxy in TNG50. Notice that while the GC mass sampling and assignment is performed at infall for all galaxies, this plot shows present-day results for surviving satellite galaxies, which means individual distributions of GCs could have been affected by tidal stripping. The top panel shows a significant drop in the average GC mass for dwarf galaxies with $M_* < 10^8 $ M$_{\odot}$, which in our model is attributed to the upper mass cut-off to sample GC mass (set to $1/100 M_{*,\rm inf}$ for each galaxy), and confirms the importance of taking this into account when dealing with GC content in low-mass dwarf galaxies \citep{forbes2018}. 

The bottom panel of Fig.~\ref{fig:mass_func} shows in cyan symbols the dispersion in the z-band magnitudes of simulated GCs associated to each galaxy, $\sigma_z$, and how it compares to the one measured in observations \citep[orange shaded region][]{jordan2007}.
Note that this relation is an extrapolation below $M_* < 10^8$ \msun. While the overall agreement is good, there is a flattening in $\sigma_z$ for our tagged GCs in high-mass galaxies, which we attribute to our absolute upper limit in the z-band luminosity/mass of individual GCs corresponding to $5\times 10^{6}$ M$_{\odot}$.

\begin{figure*}
    \centering
    \begin{minipage}{0.45\textwidth}
    \includegraphics[width = \columnwidth]{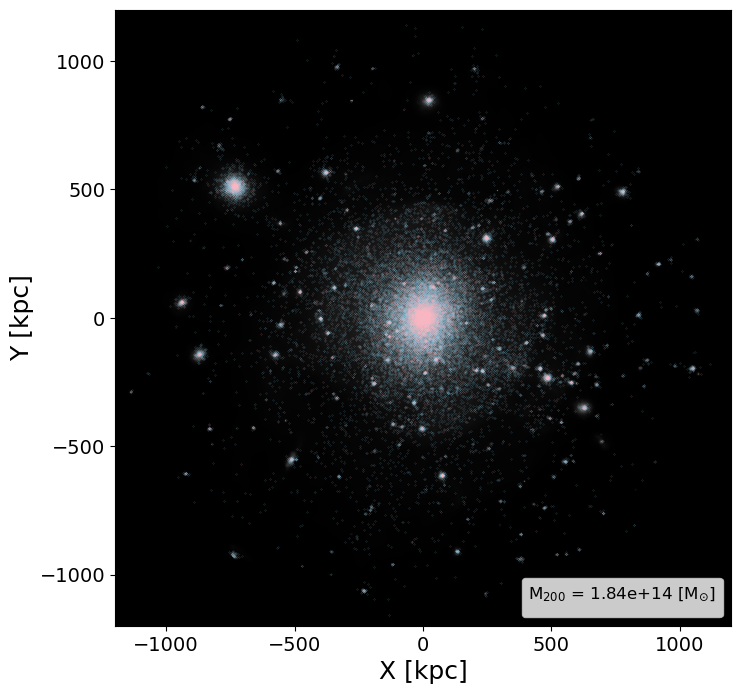}
    \end{minipage}
    \begin{minipage}{0.45\textwidth}
    \includegraphics[width = \columnwidth]{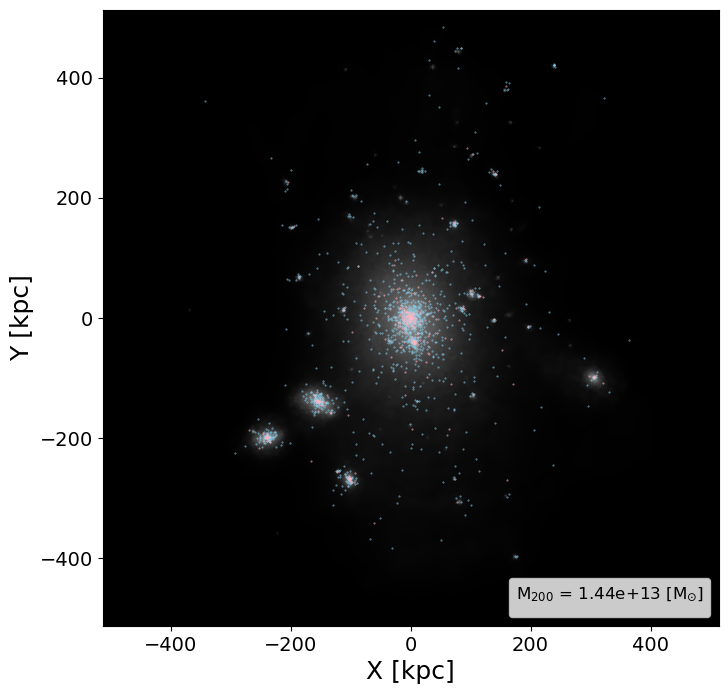}
    \end{minipage}
    \begin{minipage}{0.45\textwidth}
    \includegraphics[width = \columnwidth]{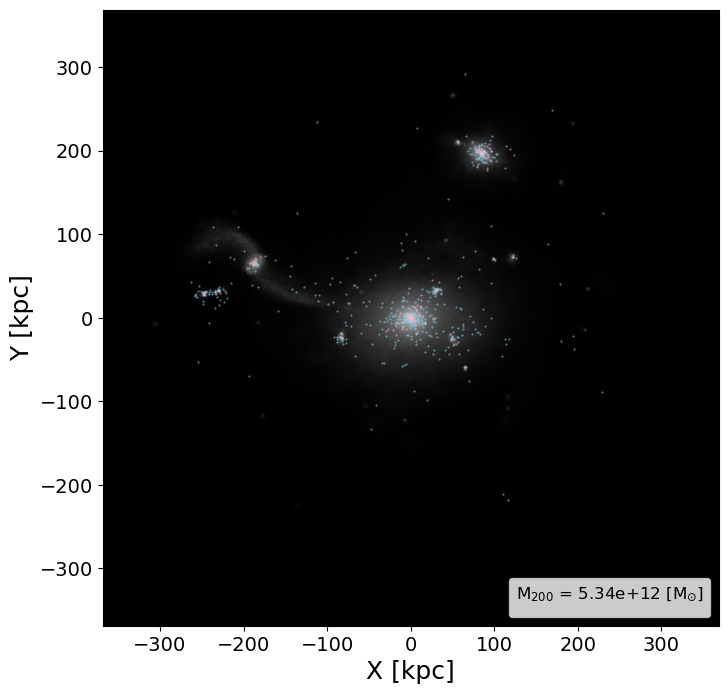}
    \end{minipage}
    \begin{minipage}{0.45\textwidth}
    \includegraphics[width = \columnwidth]{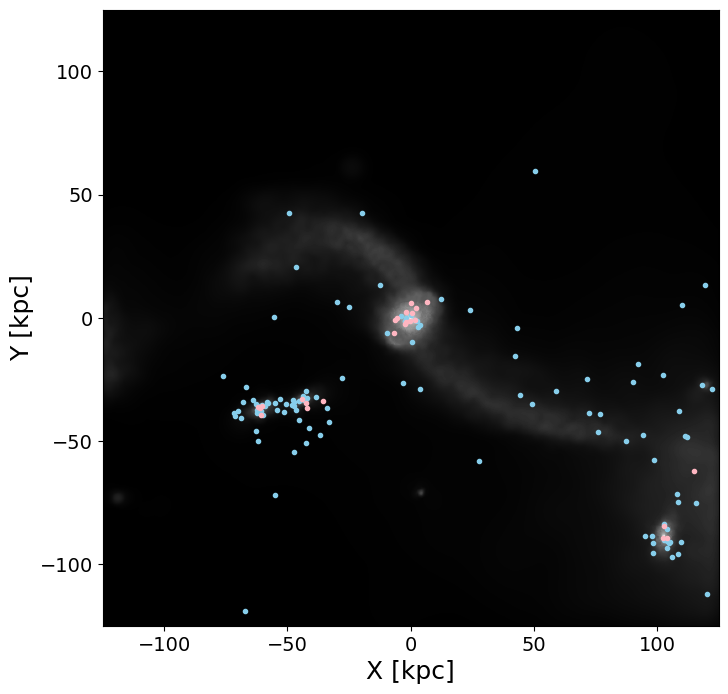}
    \end{minipage}
    \caption{Spatial maps of one realization of our GC catalog (pink and light blue points) overplotted on a visualization of the stellar density (background grayscale) for the most massive galaxy group (a Virgo or Hydra 1 analog) (top left), a galaxy group with a viral mass around $1.5\times 10^{13} $ M$_{\odot}$ (top right), and one low-mass galaxy group with a virial mass $\sim 5\times 10^{12} $ M$_{\odot}$ (bottom left). The bottom right shows a zoom-in of the GC particles associated with the stellar stream in the bottom left image. We find that the GCs distribute as expected, with the red population more spatially concentrated about their hosts and the blue component more spatially extended. We also find the presence of intracluster GCs, see Section \ref{sec:icgc} for a more detailed discussion.}
    \label{fig:projections}
\end{figure*}

We showcase some examples of our GC catalog with the final tagging results in Fig \ref{fig:projections}. Pink and light blue dots indicate our tagged realistic GCs overplotted onto the stellar density predicted by TNG50, shown in the background grayscale. To create some intuition on the range of simulated objects included in our sample, we show several systems on different mass scales, from a Virgo-like galaxy cluster in the top left of the figure, a Fornax mass galaxy cluster in the top right, and a Centaurus A mass group in the bottom left. Interestingly, it is not uncommon to find substructures of GCs in our catalog: the bottom right panel shows a set of simulated GCs that appear to be following a tidal stream in the stellar component of a disrupting host galaxy. 
We also see correct behavior of the GCs as a whole -- the red GCs are more spatially concentrated around their host galaxies than the blue GCs. \new{While this is partially imposed by design in the model, the more centrally-concentrated tagging for the red component is done at infall, while Fig.~\ref{fig:projections} shows that it is mostly preserved until $z=0$ despite tidal stripping events and interactions with the host environment.} In agreement with previous version of this tagging technique \citep{ramosalmendares2018, ramosalmendares2020}, the model predicts the formation of an intra-cluster GC component, or GCs that exist in the space between the galaxies, which is built mostly from the disruption and merging of early accreted satellite galaxies, a topic that we return to in Section \ref{sec:icgc}.

\section{Build up of the intracluster GC component}
\label{sec:icgc}

\begin{figure*}
    \centering
    \begin{minipage}{0.45\textwidth}
    \includegraphics[width = \columnwidth]{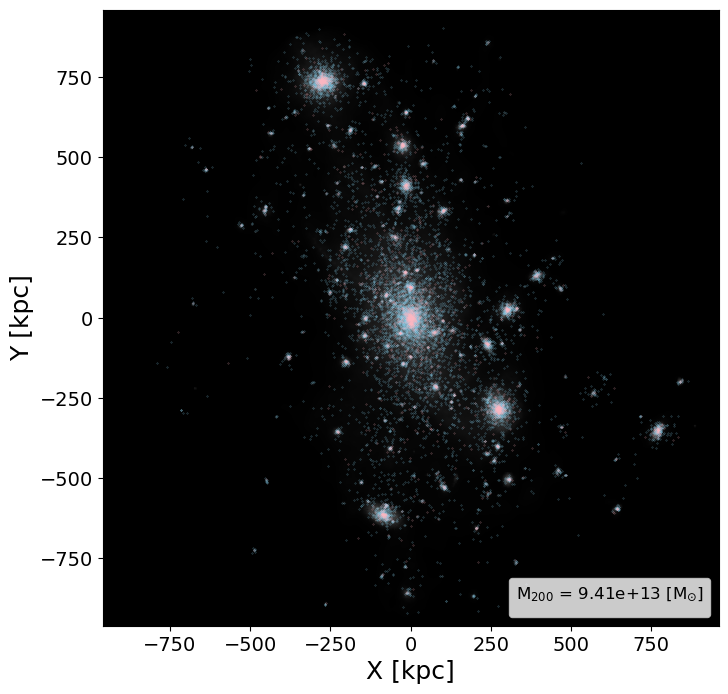}
    \end{minipage}
    \begin{minipage}{0.45\textwidth}
    \includegraphics[width = \columnwidth]{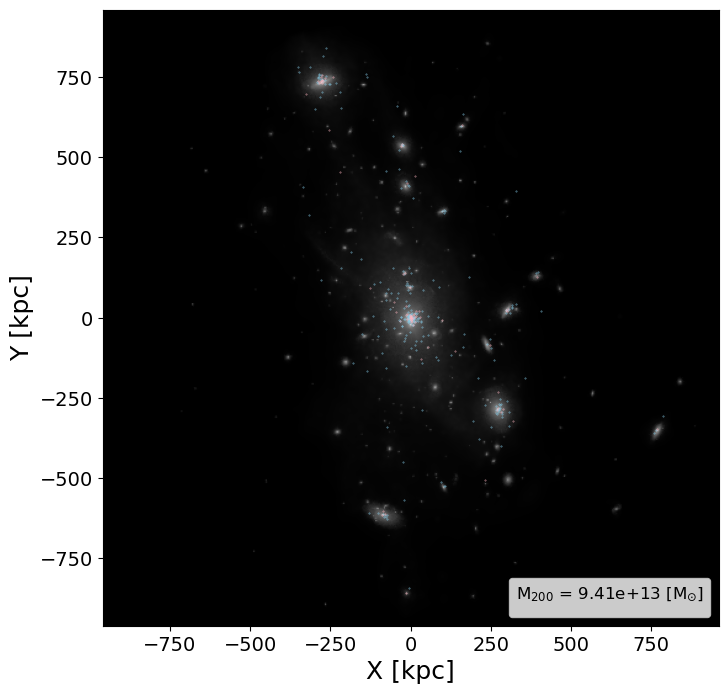}
    \end{minipage}

    \begin{minipage}{0.45\textwidth}
    \includegraphics[width = \columnwidth]{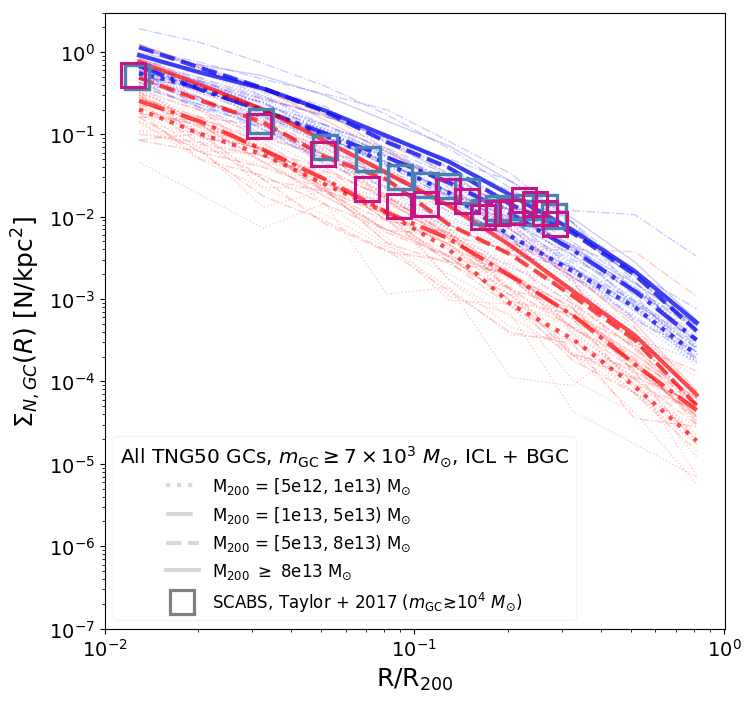}
    \end{minipage}
    \begin{minipage}{0.45\textwidth}
    \includegraphics[width = \columnwidth]{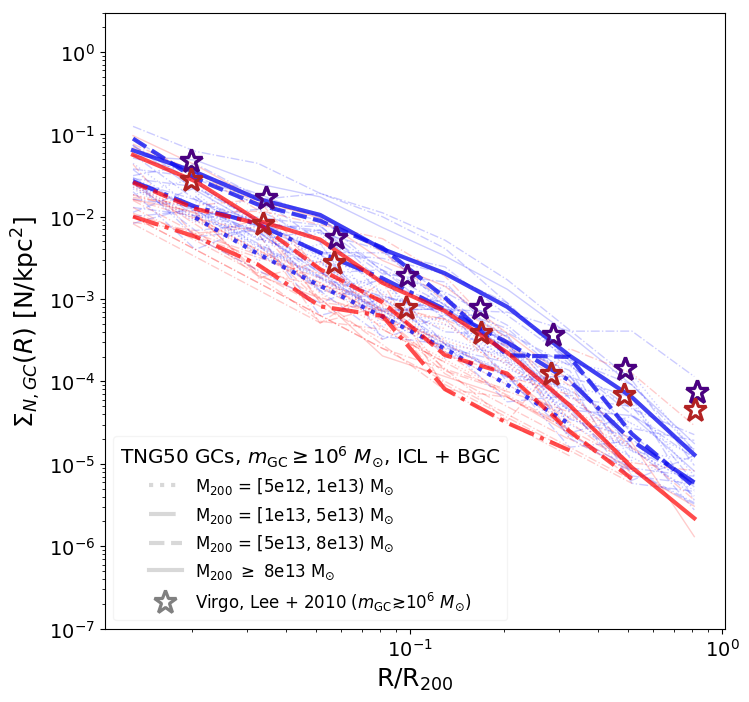}
    \end{minipage}
    \caption{Top: Projections of the stars (background grayscale) and GCs (pink and skyblue points) for FoF group 1 all the realistic GC particles associated to the group (left) and the most massive (and thus the brightest) GC particles, defined to be those with individual GC mass $m_{\rm GC} \geq 1\times 10^6$ M$_{\odot}$ (right panel). Bottom: Radial surface number density profiles for the GCs for all groups (low transparency curves) and medians for various mass bins (high alpha curves) compared to observations for the Virgo ICGCs from \citet[][red and blue stars,]{lee2010} and SCABS \citep[pink and cobalt squares, ][]{taylor2017}. The bottom left shows the profiles using all realistic GC particles and the bottom right shows the profiles using only massive GC particles, as defined for the top row. We see that this mass cut puts the predictions of the model much more in-line with what is shown in the observations from \citet[][]{lee2010}. This visually illustrates the effects of brightness cuts in observations of the ICGCs.} 
    \label{fig:projections_icl}
\end{figure*}

Observationally, the presence of GCs in the intracluster regions (or ICGCs) has been detected and surveyed in several nearby galaxy groups and clusters such as Fornax \citep{bassino2003, schuberth2008}, Coma \citep{madrid2018, peng2011}, Abell 1689 \citep{alamomartinez2017}, Virgo \citep{lee2010, durrell2014, ko2017, longobardi2018} and Centaurus A \citep{taylor2017}. Similarly to GCs in the halo of the MW \citep[see e.g., ][]{Keller2020}, ICGC studies hold the promise to help unravel the accretion history of their host halos and important properties of the progenitor galaxies building the intracluster light of the groups and clusters \citep[e.g., ][]{Villaume2020,Ko2022}.

One of the predictions of our GC model is the formation of such an accreted ICGC component, built from a combination of GCs previously associated with galaxies that have merged to the group or cluster host and also from the stripping of surviving satellite galaxies. Such a component is not directly ``tagged" or calibrated for in our simulations, but instead is the result of the hierarchical assembly of structures in $\Lambda$CDM. More specifically, while some GCs are tagged to the central galaxies in each group, this occurs when they reach a very small fraction of their final virial mass ($5\%$, see Sec.~\ref{sec:methods} for details), resulting in those GCs assigned to the central galaxy being largely subdominant (about a $\sim$dex less in GC numbers) compared to the accreted ICGCs acquired from tidal stripping and merging of the satellite galaxies. The study of the ICGC component is therefore an important benchmark of our GC model.

In this work, we define ICGCs to be GCs within the virial radius of a group or cluster host that are not currently gravitationally associated to any satellite as measured using Subfind. We note that while this differs from observational methods of determining GC membership to the ICL, which includes fitting profiles to distinguish the ICGCs from the GCs associated with the BCG \citep[e.g., ][]{taylor2017}, employing radial cuts to remove the contribution of GCs of satellite galaxies \citep[e.g., ][]{lee2010}, or using kinematic data of GCs when available \citep{longobardi2018}, this is a definition that is best physically motivated for our purposes. We have explicitly checked that using different radius cuts for satellite galaxies to distinguish between the ICGCs from the GCs of satellites, as done in some observational studies, does not substantially change the properties of the ICGCs reported here.

The top row of Fig.~\ref{fig:projections_icl} shows projections of GCs (associated to galaxies and part of the ICGCs) tagged in the second most massive group in our sample (FoF group 1), with $M_{200} \sim 9 \times 10^{13}$ M$_{\odot}$, comparable to the Virgo or Fornax clusters. As before, pink and light blue dots correspond to tagged red or blue GCs and the gray scale indicates the stellar component. Because GCs are now assigned individual masses (see Sec.~\ref{ssec:gcmass}), we can create different maps mimicking different luminosity (or mass) cuts: the left panel shows all tagged GCs in FoF 1 (or equivalent all GCs above a mass cut $7 \times 10^3$ \msun) while the right panel illustrates what would be observed in a shallower survey only able to map GCs more massive than $\geq 10^6$ \msun. 

As expected, the number of GCs decreases in the right panel due to the lower availability of more rare massive GCs. Interestingly, the substructure mapping should be different between these two images, as more massive GCs are preferentially formed in more massive galaxies (see top panel of Fig.~\ref{fig:mass_func}), leaving dwarf galaxies underrepresented in the right panel compared to the left. The extension of our model to include the masses of individual GCs makes the current GC catalog especially useful for exploring how completeness and magnitude limits might impact observational results. 

We quantify the ICGCs via their projected number density profile as a function of projected radius (normalized to the virial radius of the host) in the bottom panels of Fig.~\ref{fig:projections_icl}. The left and right panels correspond again to all GCs and GCs more massive than $10^6$ \msun, respectively. Individual thin lines (red or blue to refer to the red or blue GCs) indicate the projected radial profiles in each of our $39$ groups, while thick curves show the resulting medians when splitting our sample in four virial mass bins roughly consistent with: Virgo mass objects ($M_{\mathrm{vir}} \geq 8\times10^{13}$ M$_{\odot}$), Fornax mass objects ( $5\times 10^{13}$ M$_{\odot} \leq M_{\mathrm{200}} < 8\times 10^{13}$ M$_{\odot}$), higher-end mass estimates of Centaurus A ($1 \times 10^{13}$ M$_{\odot} \leq M_{\mathrm{200}} < 5 \times 10^{13}$ M$_{\odot}$), and lower-end halo mass estimates of Centaurus A as well as massive elliptical systems ($5\times 10^{12} \leq M_{\mathrm{vir}} < 1\times 10^{13}$ M$_{\odot}$).

There is a weak dependence of the ICGC number density on host mass, with smaller mass systems having lower number densities, but the object to object scatter is large. The red ICGCs have a slightly steeper radial distribution than the blue one, as expected from the differential stripping due to their initially more biased distribution towards the centers of their host galaxies at infall, but the effect is rather small. 

Global GC surveys are very challenging observationally for external and distant systems. As a result the available data is scarce. We compare our predictions with two available constraints: GCs in Cen A from the SCABS survey that correspond to a minimum GC mass of $\sim 10^4$ M$_{\odot}$\citep[][ red and blue squares]{taylor2017} and GCs in the Virgo cluster from \citet{lee2010}. In the bottom left panel, we see that our model shows an overall good agreement with measurements in Cen A, although we predict a steeper red ICGC component than the SCABS results. Here we are assuming a virial mass $M_{200}=10^{13}$ \msun, which corresponds to $R_{200} \sim 450$ kpc. The flattening observed in Cen A beyond $R \sim 0.2 R_{200}$  might be associated with the ring-like structure detected in this system \citep{taylor2017} and might not necessarily be present in our sample, although we do find some interesting cases where simulations also predict a flattening. We defer this study to future work. 

The observational data in Virgo corresponds to a brightness cut of $21.3$ mag in the $i$-band, which means that only the brightest $\sim 13\%$  of the GCs in the Virgo cluster are detected \citep{lee2010}. We therefore show in the bottom right panel of Fig.~\ref{fig:projections_icl} the number density profiles of GCs more massive than $10^6$ \msun, which is a better match to the shallower GC survey in Virgo using SDSS data. Here we assume a $1700$ kpc virial radius, which corresponds to a virial mass $\sim 5 \times 10^{14}$ \msun\; following \citet{Kashibadze2020}. We find a good agreement in normalization and slope of our simulated GC catalog and these observations in Virgo, with the differentiation between blue and red GCs improved with respect to \citet{ramosalmendares2020}, mostly driven by the improved numerical resolution in our simulations.

While the overall objective is not to reproduce in detail the observations of individual systems, it is reassuring to see that the predictions of our GC tagging method for ICGCs number densities are well in the ballpark of the observations available to date. This is particularly important given that this component is not directly tagged in the simulations, but instead is naturally built by the assembly process of groups and clusters. A more detailed study of the ICGC component and its relation with the build up of the intracluster light will be presented in future work (Ahvazi et al., {\it in prep}).

%
\begin{figure*}
    \begin{center}
    \includegraphics[width = \columnwidth]{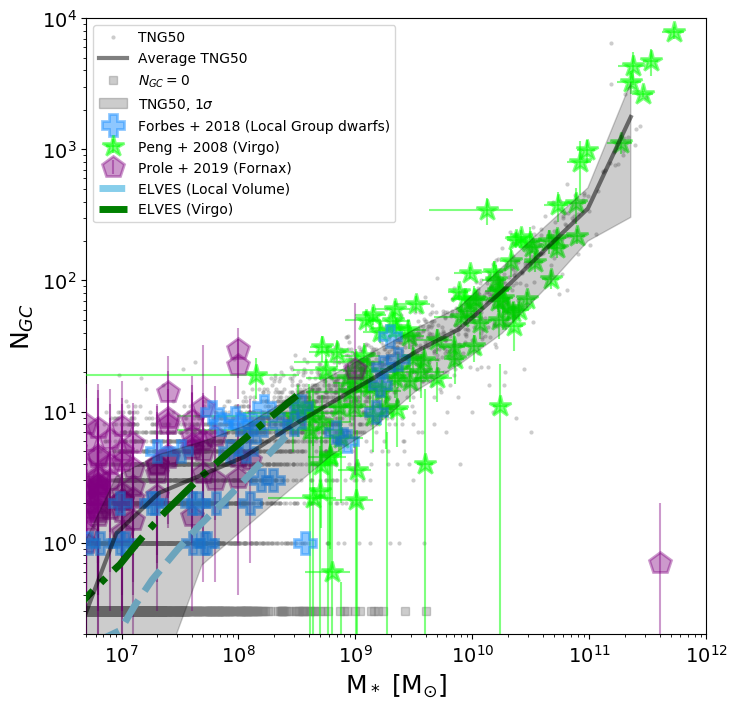}
        \includegraphics[width = \columnwidth]{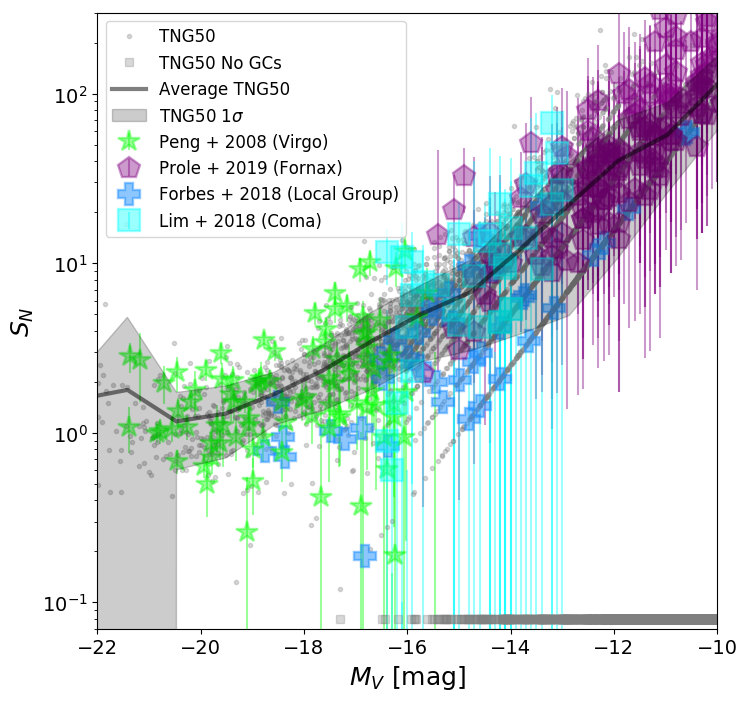}
    \caption{\textit{Left:} Number of GCs, $N_{GC}$, as a function of host galaxy stellar mass. Simulation points are plotted as gray dots, with the average shown as the solid black line and $1\sigma$ variation shown as the gray shaded region. Simulated galaxies with no GCs after the mass function selection are shown as gray squares at $N_{GC} = 0.2$. Observational data for observed galaxies are plotted as solid lime green stars \citep[Virgo,][]{peng2008}, purple pentagons \citep[Fornax,][]{prole2019}, and blue crosses \citep[Local Group,][]{forbes2018}. Average results from ELVES-II in Virgo- and Local Volume-like environments are shown in dark green dot dashed and light blue dashed lines respectively. The simulation points tend to follow the trend and scatter of the observational data. \textit{Right:} The specific frequency, $S_N$, as a function of host galaxy V-band magnitude, $M_V$. $M_{V}$ and $S_N$ have been corrected to correspond to the mass to light ratio observed for Virgo (see Appendix \ref{app:colors}). Colored shapes correspond to the same observations as before, with the addition of cyan squares \citep[Coma][]{lim2018}. Galaxies with $S_N = 0$ are shown as gray squares at $S_N = 0.07$. The agreement of both measures of GC abundance with observations in terms of shape and scatter suggests that the assumption that GC mass scales with halo mass holds to a reasonable extent, even into the dwarf regime.}
    \label{fig:ngc}
    \end{center}
\end{figure*}

\section{GC content in dwarfs to giant galaxies}
\label{sec:GCs_in_galaxies}

The GC tagging model calibrates the total mass of globular cluster systems in galaxies at $z=0$ using the $M_{\rm GC}-M_{200}$ power-law relation from \citet{harris2015}. As explained in detail in Sec.~\ref{ssec:tag}, only simulated halos with calculated $M_{200} > 10^{11}$ \msun\; participate in the calibration, while lower-mass objects are assumed to follow an extrapolation of that power-law. The GC content of dwarf galaxies in halos less massive than $M_{200} = 10^{11}$ \msun, corresponding to $M_* \sim 10^9$ \msun\; in the stellar-halo mass relation of TNG50, is therefore a prediction of the model under this assumption. We explore in this section how the results obtained in the regime of dwarf galaxies compare to current observational constraints. 

\subsection{Number and Specific Frequency of GCs}
\label{sec:N_GCs}

Fig. \ref{fig:ngc} shows in the left panel the relation between the number of globular clusters, $N_{\rm GC}$, and the host galaxy stellar mass, $M_{*}$, in our simulated systems (gray symbols). Individual galaxies are shown in gray points,  with the median relation (including galaxies with $N_{\mathrm{GC}} = 0$) shown as the solid black line, with the $25\% - 75\%$ dispersion shown as the gray shaded region. Our results agree well in both overall shape and dispersion with available constraints from observations  shown here in colored symbols: green stars from galaxies in the Virgo cluster \citep{peng2008}, magenta pentagons for dwarfs in the Fornax cluster \citep{prole2019}, and additional low-mass galaxies from the Local Volume in sky-blue crosses \citep{forbes2018}. We also indicate the average results from ELVES-II reported in \citet{Carlsten2022}, showing a low-mass selection of Virgo cluster dwarfs (green dot-dashed curve) and dwarfs in the Local Volume (dashed light blue). While our sample does not include low density environments such as the Local Volume, our average values for the lowest mass objects resolved in our sample track well the slope of the average number of GCs per system observed in ELVES-II. 

For completeness, we also show the related quantity, specific frequency of GCs or $S_N$, as a function of V-band magnitude in the right panel of Fig.~\ref{fig:ngc}. We calculate the specific frequency following \citet{harris1981}:

\begin{equation}
    S_N = N_{\mathrm{GC}} 10^{0.4(M_V + 15)} ,
\end{equation}

\noindent
where $N_{\mathrm{GC}}$ is number of GCs and $M_V$ is the V-band absolute magnitude of the host galaxy. For most galaxies we take $M_V$ directly from the simulation, except for the high-mass galaxies ($M_* > 10^{9} M_\odot$), where we adopt a fixed mass-to-light ratio equal to $3.6$ to convert from mass to luminosity following observations in the Virgo cluster \citep{peng2008}.

The color coding on the right panel of Fig.~\ref{fig:ngc} is the same as introduced for the left panel, with our simulated galaxies shown in gray and a set of available observational constraints using color symbols with error bars. We show galaxies with $S_N = 0$ as gray squares with $S_N = 0.07$ so that they are visible on the log scale. The median and $25$-$75$ percentiles are calculating not including galaxies with $S_N =0$. 

Simulated $S_N$ values overlap well with observational constraints, 
 in particular in the regime of dwarf galaxies, where typical $S_N$ values of several dozens to a few hundreds become common for dwarfs fainter than $M_V \sim -13$. The inclined lines seen for simulated galaxies with $M_V > -16$ correspond to discrete numbers of GCs (galaxies with $1$, $2$, $3$ GCs) and seem to represent well several of the dwarf galaxies in the \citet{forbes2018} sample. 

While, to a certain degree, the agreement in the high-mass end of Fig.~\ref{fig:ngc} might be expected because of the calibration of our model to follow the $M_{\rm GC}$-$M_{200}$ relation, it is not fully guaranteed due to the following factors: $(i)$ our method tags the satellite population at infall and not at $z=0$, $(ii)$ we tag based on halo mass and not $M_*$ as shown here where galaxies continue to evolve their $M_*$ and $M_V$ after infall and $(iii)$ we tag on total GC mass, $M_{\rm GC}$, not specifically in GC number. Most importantly, our simulations compare well with measurement of GC numbers in dwarf systems below those used to calibrate the $M_{\rm GC}$-$M_{200}$ relation, offering support to the hypothesis that this power-law relation between GC mass and halo mass extends at least to objects with $M_* \sim 5 \times 10^6$ \msun.

Interestingly, the left panel of Fig.~\ref{fig:ngc} shows that the average number of GCs continues to decrease with smaller $M_*$ in the full range explored here (when including zeros). This is relevant because it helps rule out \new{more extreme,} ``purely stochastic" models where the number of GCs is simply a random number in the low-mass end
\citep[e.g., ][]{El-Badry2019}. \new{We note that this purely stochastic model is not proposed as physically-motivated, but instead used in \citet{El-Badry2019} as an interesting extreme behaviour to explore the slope of the relation between halo and GC mass}. Such purely stochastic models, while being able to reproduce the high-mass end of the power-law relation $M_{\rm GC}$-$M_{200}$ due to mergers and hierarchical assembly, would provide a much shallower or constant average number of GCs with $M_*$ in the low-mass regime where stochasticity starts to dominate. Instead, our results agree well with the conclusions presented in \citet{forbes2018}, where the slope and scatter of the GC content is consistent with a model where dwarf halos lay on an extrapolation of the GC - halo mass relation measured for more massive systems. 

\subsection{Radial extent of GCs}
\label{sec:R_GCs}

\begin{figure}
    \centering

    \includegraphics[width = \columnwidth]{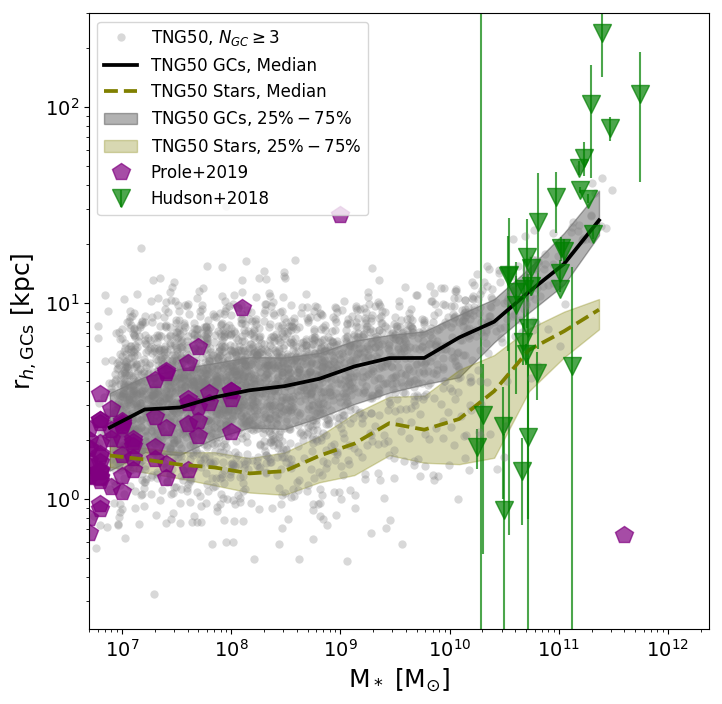}

    \includegraphics[width = \columnwidth]{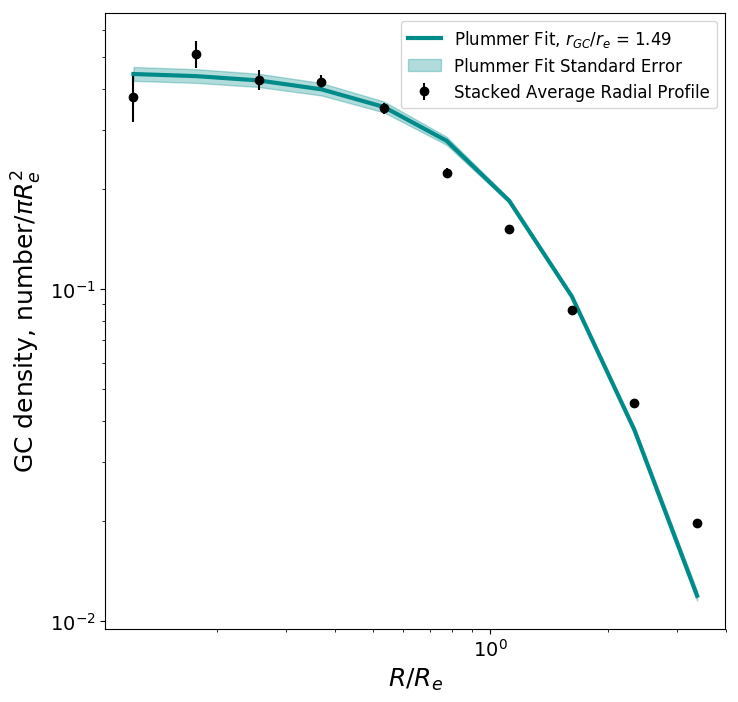}
    \caption{\textit{Top:} The half number radius of GCs around TNG50 galaxies (gray points), low surface brightness galaxies from \citet{prole2019} (purple pentagons), and higher-mass galaxies from \citet{hudson2018} (green triangles). We show simulated galaxies with $N_{\rm GC} \geq 3$. We find good agreement with observations for higher-mass galaxies, but find a flatter slope than the observations for dwarf mass galaxies. We do note however that the scatter in the simulated points covers the range seen in observations. \textit{Bottom:} Low number of GCs in dwarfs may favour using alternative methods to measure a half-number radii than individual counts. Inspired by observational methods of calculating $R_{h, \mathrm{GCS}}$ in dwarfs, we show the average {\it stacked} radial profiles for the GCs of dwarf galaxies with stellar masses between $5\times 10^6$ M$_{\odot}$ and $10^{8.5}$ M$_{\odot}$. The solid dark cyan line shows the best fitting Plummer profile for the stacked GCs. The error bars and the shaded region are obtained via bootstrapping. Our best-fit profile recovers the expected factor of $\sim 1.5$ that relates the stellar half light radius, $R_e$, with the GC half number radius, $R_{\mathrm{GC}}$, which is consistent with observational estimates.}
    \label{fig:rhalf}
\end{figure}

We show in Figure \ref{fig:rhalf} our predictions for the (3D) radial extent of the tagged GC systems as a function of stellar mass. We use the half-number radius $r_{h,\mathrm{GC}}$ to characterize the radial extent of the GC systems, which we calculate by rank-ordering the GCs associated to each galaxy in increasing distance to their host and finding the radius of the GC that divides the sample in two. It is expected that the accuracy of this estimate scales with the number of GCs, with dwarf galaxies having the largest uncertainties given their low number of GCs. In this figure, we include only simulated galaxies with $N_{\rm GC} \geq 3$ (gray circles), which allows for the determination of  $r_{h,\mathrm{GC}}$ (this cut in $N_{\rm GC}$ might not necessarily apply in observations (purple pentagons and green triangles), where the half number radius is determined via profile-fitting, see below). Projected sizes in observations have been converted to 3D by multiplying the reported values by a $(4/3)$ factor, which assumes a spherical distribution \citep{Somerville2018}.

\new{Given the relatively high spatial resolution of TNG50 ($\sim 290$ pc at $z = 0$), the radial extents of the GC systems considered here are numerically well resolved. Their typical sizes} increase from a few kpc for dwarfs with $M_* \sim 10^7$ \msun\; to $r_{h, \mathrm{GC}} \sim 40$ kpc for our largest satellite galaxy with $M_* \sim 10^{11}$ \msun, with a significant object-to-object scatter, in particular at the low-mass end. The median trend is highlighted by the black solid line, with shaded regions indicating the $25$-$75$ percentiles in our sample. In agreement with observations, simulated GCs are typically more extended than the stellar component in galaxies, which is indicated by the gray-green shaded curve and shaded area showing the median and $25$-$75$ percentiles of the half-mass radius of the stars in the same galaxies. On average, GCs are a factor $\sim 2$-$3$ times more extended than the stars in galaxies, with a hint at a smaller ratio for low-mass galaxies. 

We reproduce well typical sizes for GC systems in MW-mass galaxies, predicting $r_{h, \mathrm{GC}} \sim 10$ kpc for galaxies with $M_* = 5 \times 10^{10}$ \msun, and an increasing size with mass, in good agreement with data from \citet{hudson2018}. This is not completely surprising since the scale parameters in the Hernquist profiles used to tag the red and blue GC components at infall were partially chosen in the original model \citep[see ][]{ramosalmendares2020} to reproduce typical GC distributions in these scales. 

It is interesting, however, to explore what predictions arise from extrapolating the same scaling towards low-mass galaxies. We show with purple pentagons data from dwarfs in the Fornax cluster, taken from \citet{prole2019}. While our systems overlap with the observed dwarfs, simulated galaxies seem to have systematically larger half-number radii than observations. However, we caution that the exact size measured is very sensitive to the definition chosen in systems dominated by low-number statistics, like GCs in dwarfs.  

The bottom panel of Fig.~\ref{fig:rhalf} shows a different approach, often used in observations of dwarf galaxies: determining the size of GCs based on profile-fitting of the resulting stacked GC profile (instead of individual GC counting in each galaxy as in the upper panel). We show the stacked projected number density profile of GCs for dwarf galaxies with $5 \times 10^6 < M/M_\odot < 10^{8.5}$ as a function of (projected) radius normalized to the effective radius for each dwarf \citep[see for instance ][]{Carlsten2022}. Errorbars and the shaded region are calculated via bootstrapping and correspond to the r.m.s values from those realizations.
The dark cyan line shows the best fitting Plummer profile for our simulated dwarfs, which suggests that the half number radius of GCs in these systems is $\sim 1.5$ times the half-mass radius\footnote{We have explicitly checked that on the high mass end, provided that the chosen profile provides a good fit to the individual GC distributions, the half mass radius computed via GC counting or via profile fitting are within statistical uncertainty of each other, being therefore less of an issue for massive galaxies with a numerous GC population than in low mass dwarfs with only a few GCs.}, which is in good agreement with observational estimates \citep{georgiev2010, Carlsten2022}.

Since in our model, the scaling of the GC tagging depends only on the dark matter half-mass radius (through the calculation of the best-fitting NFW profile at infall), the good agreement with the scaling of GCs and the {\it stellar}  component of galaxies is, again, not guaranteed and an interesting feature of our catalog. It also points to another puzzling link between GCs and dark matter halos, in this case through radial extent instead of total mass, that may shed light on the origin and formation of GCs. 

\subsection{Dependence on environment} 

Recently, \citet{Carlsten2022} reported a higher GC content for dwarf galaxies in the environment of Virgo compared to dwarf satellites of the same mass in lower density environments of the Local Volume. This finding has been interpreted as an extension of a radial trend in the Virgo and Coma cluster where dwarf galaxies near the center (and therefore on higher density regions) have on average a larger specific frequency compared to those located further out \citep{peng2008, lim2018}. Such a trend has been explained as a natural consequence of dwarfs with inner orbits and higher environmental densities having formed their stars earlier on, with more intense star formation histories leading to the formation of GCs with a higher specific frequency \citep{peng2008,mistani2016} than objects in the field. 

Additionally, since dwarfs stopped forming stars in high density environments earlier than those in the field, comparing them at fixed $M_*$ today means that the dark matter halos of those in high density environments are biased high. This follows since quiescent dwarfs today should have continued forming stars reaching higher luminosities at the present day had they stayed in the field \citep{mistani2016}. Such an effect would also lead to a higher GC-content for early type dwarfs in groups and clusters.  

Our GC catalogs sample a relatively narrow range of environments, including groups and low-mass clusters with $M_{200}=[5 \times 10^{12} \rm - 2 \times 10^{14}]$ \msun\; and no dwarfs around MW-type galaxies (such as those in the low density regions of ELVES) or directly in the field. However, we have explicitly checked that, within the range of environments of our sample, we find no significant difference in the predicted GC number or specific frequency for simulated dwarfs in low-mass vs. high-mass host halos, nor do we find a trend with cluster-centric radii.

Our method is unable to link the GCs to the star formation histories (only infall virial mass is used to tag the GCs onto our galaxies). However, the second effect (related to the higher halo mass for dwarf galaxies in higher density environments) is naturally taken into account in our catalog. We find no significant difference in the infall mass or infall times of the surviving dwarf population between our simulated groups, which partially explains the lack of correlation between $N_{\rm GC}$ or $S_N$ with environment seen in our sample (see Appendix \ref{app:no_environ}).

Noteworthy, in agreement with our predictions, dwarfs in the Fornax cluster also show no enhancement in GC number of specific frequency \citep{prole2019} when compared to dwarfs in the  \citet{georgiev2010} field sample. This might suggest that while the mode of star formation and differences in halo masses may imprint an excess of GCs for dwarfs in higher density environments, those effects set in at higher density environments (closer to those of Virgo and Coma clusters, $M_{200} > 5 \times 10^{14}$ \msun) than those simulated here. Surveys of dwarfs in intermediate-mass groups and low-mass clusters are needed to confirm this hypothesis and determine whether or not our GC tagging model might benefit in the future from including additional GC formation channels. For instance, an increased number of GCs forming in starburst events associated with pericenter passages have been shown successful at explaining a cluster-centric radial gradient in GC content for cluster dwarf galaxies \citep[e.g., ][]{mistani2016} and the higher GC content in ultra-diffuse galaxies \citep[e.g. ][]{carleton2021}.

\begin{figure}
    \centering

    \includegraphics[width = \columnwidth]{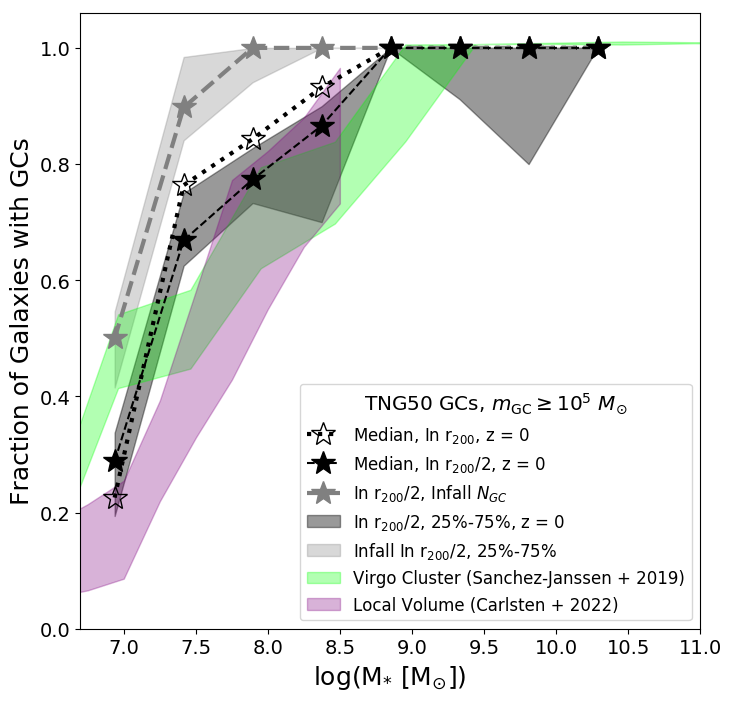}
    \caption{A look at GC occupation fraction (defined as the fraction of galaxies that have at least one GC associated to them) as a function of stellar mass for galaxies within $r_{200}/2$ of our selected groups. The figure shows the occupation fraction from TNG50 by infall number of GCs as the gray dotted line with stars (with the gray shaded region showing the $25\% - 75\%$ spread between environments); z = 0 values are shown within $r_{200}$ by the unfilled black stars and dotted line and within $r_{200}/2$ as the filled black stars and dashed line (with the $25\% - 75\%$ spread between environments shown as the black shaded region). Observed occupation fractions from Virgo \citep{sanchezjanessen2019} and the Local Volume \citep{Carlsten2022} are shown as lime green and magenta shaded regions respectively. The difference between the dim gray and the black filled stars shows that tidal stripping has a sizable effect on setting the occupation fraction in dwarfs with $M_* < 10^9$ \msun. Lower-mass dwarfs with $M_*<10^{7.5}$ \msun\; have additionally a 50\% occupation fraction already at infall, which we explain through their low total GC mass together with the stochastic sample of the GC mass function.}
    \label{fig:occ_frac}
\end{figure}

\section{GC Occupation Fraction} 
\label{sec:occupation_fraction}

While all massive galaxies appear to have associated GCs, the same is not true for low-mass galaxies, some of which are observed to host no GCs. The GC occupation fraction, defined here as the fraction of galaxies at fixed stellar mass that host at least $1$ GC, is an important constrain on GC formation scenarios and is fundamental to determine the minimum galaxy mass able to form GCs that survive until the present day. 

As discussed in Sec.~\ref{sec:intro}, observationally, the GC occupation fraction is found to be close to one for galaxies with stellar masses  $M_* \geq 10^9$ \msun, and to sharply decrease for lower-mass galaxies \citep{sanchezjanessen2019,Eadie2022,Carlsten2022}. An important caveat of these studies is that the low-mass galaxies included are mostly satellite objects, although the host mass varies from the Virgo cluster to satellites of $\sim L*$ hosts in the Local Volume. We can use our GC catalog to compare with these observations and to determine the role of tidal stripping in satellite galaxies in establishing such a trend.  

Black starred symbols in Fig \ref{fig:occ_frac} show the median z = 0 GC occupation fraction in our simulated galaxies as a function of host galaxy stellar mass for GCs with individual masses $m_{\rm GC} > 10^5$ M$_{\odot}$ to mimic the brightness cutoff from \citet{sanchezjanessen2019} in the Virgo cluster. In agreement with observations, our catalog predicts a decreasing occupation fraction for dwarfs with $M_* \leq 10^9$ \msun, while all galaxies more massive than that are expected to host GCs. We find a weak dependence of the occupation fraction with cluster-centric radius, with occupation fraction being only slightly lower when considering satellites in the inner regions of simulated groups and clusters ($r < r_{200}/2$, solid black stars, short dashed curve) compared to including all satellites within the virial radius (open black stars, dotted curve). Shaded regions indicate $25$-$75$ percentiles of our sample. 

Encouragingly, our present-day occupation fraction agrees well with available measurements. For instance,  green shaded area corresponds to galaxies within $\sim R_{200}/3 - R_{200}/2$ of the Virgo cluster \citep{sanchezjanessen2019} while the occupation fraction in satellite dwarfs within the Local Volume is shown in magenta \citep[][]{Carlsten2022}. We find little variation in occupation fraction with environment across the 39 simulated groups in TNG50, which agrees well with findings reported in the two environments explored by the ELVES survey \citep{Carlsten2022}.

For comparison, we show in gray the ``initial" occupation fraction, e.g., the occupation fraction measured at the infall time for our simulated dwarfs within the $r_{200}/2$ sample (gray starred symbols and dashed curve for the median, shading indicating $25$-$75$ percentiles). Differences between the gray curve at infall and the black curve today is a direct measure of the impact of tidal stripping of GCs by the host groups and clusters, which seems to be substantial for dwarfs with $10^{7.5} < M_*/M_\odot < 10^9$. 

In particular, our model predicts that all dwarfs with $M_* \sim 10^8$ \msun\; should host at least one GC with mass $\sim 10^5$ \msun in the field, while such dwarfs have only $75\%$ occupation fraction on average when observed in groups and clusters. This is a testable prediction that might be confirmed or refuted when large observational samples of field dwarfs with their GCs become available. 

On the other hand, for stellar masses lower than $M_* \sim 10^8$ \msun, the prediction for the infall GC occupation fraction is already lower than $1$. For instance, our model predicts that only half of the dwarfs with $M_* \sim 10^{7}$ \msun\; hosted at least one GC with $M_* \geq 10^5$ \msun\ at infall. 

Dwarfs with  $M_* \sim 10^{7}$ \msun\; have a ``halo mass" $M_{200} \sim 7\times 10^9$ M$_{\odot}$ (calculated following abundance matching from \citet{hudson2015} as described in Sec.~\ref{ssec:tag}) and Fig.~\ref{fig:masscal} shows that for such objects the median mass in GCs is $M_{\rm GC} \sim 2.6\times 10^5$ \msun. This seems above our GC mass $=10^5$ \msun considered for this occupation fraction calculation, raising the question of why the occupation fraction is lower than $1$ at infall. 

We find that the scatter around the $M_{\rm GC}$-$M_{200}$ relation coupled to the stochastic sampling of the GC mass function (see Sec.~\ref{ssec:gcmass}) makes the chances for dwarfs of this mass to host a GC with $M=10^5$ \msun\; about half. Indeed, the maximum GC mass in our model is set to be one hundredth of the mass of the dwarf (limited inspired by observations of dwarfs in the Local Group), placing a GC with mass $10^5$ \msun\ close to the upper limit of the mass distribution and therefore relatively unlikely from a random normal draw (see purple histogram on the upper panel of Fig.~\ref{fig:mass_func} for the typical mass function of GCs in this mass range). 

\new{We have explicitly checked that removing the $0.01 M_*$ cut for the sampling of GCs increases the occupation fraction slightly on the low mass end while leaving it unchanged for $M_* > 10^8$\msun. For instance, in our lowest mass bin the occupation fraction increases by a factor $\sim 2$ as a result of a less restrictive mass distribution from which to draw the individual cluster masses. Given the observational uncertainties and variations between the Virgo and Local Volume measurements, our predictions for $M_* \sim 10^7$\msun\; remain consistent with observations.}

While stochasticity explains the initial low occupation fraction, we note that at present day, the occupation fraction has additionally dropped to $25\%$ for dwarfs with $M_* \sim 10^ 7$ \msun, which is due, similarly to more massive satellites, to tidal stripping from the host. This value is in good agreement with results from the Local Volume, but it is slightly lower than that measured for the Virgo cluster. Observations of dwarfs in the field for this mass range might also help constrain if our model is initially underpredicting the occupation fraction. 

Another possibility is that projection effects in high-density environments such as the Virgo cluster could artificially be increasing the occupation fraction of low-mass dwarfs by assigning GCs from the intracluster component or from neighboring galaxies to these dwarfs. Occupation fraction being a requirement of only $1$ GC is certainly subject to significant Poisson noise, which worsens in environments with a high background component such as clusters. We will use our catalog to explore projection effects in future work. 

We highlight that the numbers presented in this section should be taken as upper limits assuming no additional GC destruction mechanism is at play after the tagging time at infall. This might not always apply, in particular in cases where dynamical friction timescales might be short, for instance, for low-mass galaxies. We show in Appendix~\ref{app:dynamical_friction} that considering the effects of dynamical friction does not significantly change our results. We conclude that the occupation fraction predicted by our model is in reasonable agreement with current observational constraints and that additional data from other environments, and more specifically, from the field, would help verify (or reject) the predictions of our model.

\section{Summary} 
\label{sec:summary}

In this work, we present a catalog of GCs tagged to the $39$ most massive groups in the TNG50 simulation. Our systems have virial masses in the range $M_{200}=[5 \times 10^{12}$ - $2 \times 10^{14}]$ \msun\; providing simulated analogs of massive ellipticals in the field to low-mass galaxy clusters. Known systems in this range may include Cen A, Fornax, Hydra-I or the Virgo cluster, where GC data is abundant. Our GC tagging technique follows from the one already applied to galaxy clusters with $M_{200} \geq
10^{14}$ \msun\; in the Illustris simulation \citep{ramosalmendares2020}, with improvements to take full advantage of the increase in resolution and the inclusion of lower-mass dwarfs in our sample. 

Briefly, our GCs are tagged to any satellite galaxy identified in the merger tree to have a maximum stellar mass $M_{*,\rm max} \geq 5 \times 10^6$ \msun\; and that has ever interacted with our host groups. For each satellite, we identify dark matter particles in its subhalo at infall with a given energy distribution that is consistent with the phase-space that we choose for the GC systems. All galaxies are tagged at infall, after which the dynamics of their assigned GCs is followed by the simulation until the present day. This enables the prediction of GC content in galaxies from dwarfs to giant ellipticals with stellar masses in the range: $M_*=[5 \times 10^6 \rm - 6 \times 10^{11}]$ \msun. GCs are tagged to more than $8000$ simulated  galaxies across time, of which more than $5000$ survive in our sample at $z=0$. 

We include a new modeling of the GC mass function that allows us to assign individual GC mass to each tagged particle. This is a necessary improvement over the previous model in \citet{ramosalmendares2020} which assigns all tagged particles equal GC mass. As discussed in Sec.~\ref{ssec:gcmass} and \ref{sec:N_GCs}, this addition is fundamental to reproducing the GC content in dwarf galaxies. 

The GC tagging method relies on only one strong assumption: galaxies at infall follow a power-law relation between mass in GCs and halo mass, with a normalization and slope that is calibrated to reproduce the present-day $M_{\rm GC}$-$M_{200}$ relation from \cite{harris2015}. Most importantly, this relation is known to hold only for galaxies with stellar mass $M_* \sim 10^9$ \msun\; and above. We therefore consider only galaxies with halo mass $M_{200} > 10^{11}$ \msun\; (or equivalently, $M_* \sim 10^9$ \msun) to participate in the calibration, while applying the calibrated relation to lower-mass galaxies as well. This approach allows us to make predictions on the GC systems of dwarfs with $M_* < 10^9$ \msun\; under the assumption that they follow an extrapolation of the same power-law of more massive systems. In this paper we compare these predictions with available observational data on GCs of dwarf galaxies. Our main results can be summarized as follows.

\begin{itemize}
    \item The GC tagging method naturally gives rise to the formation of an intracluster GC (ICGC) component which is in good agreement with the currently available data. Our individual GC-mass modeling allows the construction of mock observations of GCs at different brightness/mass cutoffs, which might prove a very useful tool for theory/observation comparison once more ICGC systems are mapped in groups and clusters. 
    
    \item The predicted number ($N_{\rm GC}$) and specific frequency ($S_N$) of GCs in dwarf galaxies with $M_*=[5 \times 10^7 \rm - 10^9]$ \msun\; are consistent with observations of dwarfs in the Local Volume as well as in clusters such as Virgo and Fornax. This provides support to the idea that low-mass dwarfs lay in an extrapolation of the GC mass - halo mass relation of more massive counterparts, in agreement with conclusions from \citet{forbes2018}. In particular, the average number of GCs as a function of galaxy mass seems in agreement with that reported for the ELVES survey in low-mass objects \citep{Carlsten2022} and it is different from one where the number of GCs is simply a random draw in the low-mass end. 
 
    \item The radial distribution of GCs around satellites in a wide range of masses is also well reproduced in our catalog, with median values ranging from $r_{h, \rm GC}\sim 2$ kpc for low-mass dwarfs with $M_* \sim 10^7$ \msun\; to $\sim 25$ kpc for $M_*= 2 \times 10^{11}$ \msun. A closer inspection to the GCs in dwarf galaxies indicates that the low number of GCs expected might bias high the estimates of the half-number radius obtained by simply rank-ordering the identified GCs in distance. When stacking GCs of similar-mass dwarfs and finding a best-fit profile, as often performed in observations, we find that the half number radius of GCs in dwarfs is closely related to that of the stars, $r_{h, \rm GC} \sim 1.5 r_{h,*}$, which is a common assumption in the literature. This is substantially smaller than the factor $\sim 3$ - $5$ between GCs and the size of the stellar component in more massive galaxies like the MW and giant ellipticals.  
    
    \item We predict a steeply declining GC occupation fraction for dwarfs with $M_* < 10^9$ \msun, which is in reasonable agreement with current constraints from Virgo \citep{sanchezjanessen2019} and the Local Volume \citep{Carlsten2022}. In our model, tidal stripping plays a significant role at lowering the occupation fraction for all dwarf galaxies, and this effect cannot be neglected when interpreting occupation fraction data in observations. For instance, we predict almost $100\%$ occupation for dwarfs in the field with $M_* = 10^8$ \msun\,  e.g., hosting at least $1$ GC with stellar mass $10^5$ \msun, 
    while in group and cluster environments the fraction is $\sim 75\%$, in agreement with observations. For lower-mass dwarfs, stochasticity in the sampling of the GC mass function coupled to their low GC mass content (set by  their low halo mass) results in the expectation of only one in two dwarfs with $M_* \sim 10^7$ \msun\; hosting a $\sim 10^5$ \msun\ GC at infall.
    For comparison, tidal stripping effects lower this to one in four for the group and cluster environments.  
\end{itemize}

Our GC tagging method is linked to an empirical calibration of the GC mass - halo mass relation and does not specifically model the formation of GCs. However, some of the results might be used to shed light on GC formation mechanisms. For example, our model naturally predicts the scaling of the size of GC systems across all masses to the dark matter halo distribution (through the half-mass radius in dark matter). The fact that we find a good agreement  with observations on the typical GC system sizes from dwarfs to large galaxies suggests another puzzling link between dark matter halos and GCs, besides the scaling on mass. An interesting link between the GC sizes and the estimated virial radius has been observationally found in galaxies with mass comparable to the MW and above \citep{hudson2018}. Our results suggest that a tight link between these two radii extends all the way into the dwarfs regime. 
 
In particular, the GC radial extent in the regime of dwarf galaxies seems in agreement with the predictions from the model where GCs form at the centers of their own dark matter halos, or mini halos, as first suggested by \citet{Peebles1984}. While this is not true for more massive galaxies, where such a ``cosmological" origin of GCs would predict radial distributions that are too extended compared to MW-like galaxies \citep{Creasey2019}, in the regime of dwarfs, the hierarchical clustering of these primordial mini-halos is of order few kpc, which is in good agreement with observations and predictions of our model \citep[see Fig. 4 in ][]{Creasey2019}. 

This suggests that, if GCs can form in their own mini-halos and hierarchically assemble in the halos of galaxies today, the best sites to look for such objects might be dwarf galaxies, where a larger fraction of GCs would be consistent with  a cosmological origin. Ultimately, only measurements of individual GC ages and metallicities would be able to fully differentiate between a primordial GC formed in its own dark matter halo, from a GC formed via baryonic processes in the ISM of galaxies \citep{Bastian2020}. Targetting GCs around dwarf galaxies with $M_* \sim 10^7$-$10^8$ \msun\; might give us the best opportunity to narrow down GC origins.

More broadly, the GC catalog presented in this work is a useful resource to study the 6D properties of GCs in groups and clusters, environments where the ab initio formation of GCs in cosmological simulations is not yet feasible. Targeting $\sim 40$ systems allows the study of halo-to-halo variations and the understanding of the link between GC properties and particular assembly history of each group; a goal that we will pursue in  future work. The GC catalog created herein is made publicly available as part of the IllustrisTNG data release.

\section*{Acknowledgements}
We would like to thank the referee, Oleg Gnedin, for a constructive and useful report that helped improved an earlier version of this manuscript. The authors would like to thank Chervin Laporte and Felipe Ramos-Almendares for useful discussions in the early phases of this project and Volker Springel for guidance and early access to TNG50 results.
JED and LVS are grateful for financial support from the NSF-CAREER-1945310 and NASA ATP-80NSSC20K0566 grants. JED also acknowledges support to the DAAD through their short term research grant and hospitality from research and administrative staff members at the Max Planck for Astrophysics during a 3-months exchange visit.
DN acknowledges funding from the Deutsche Forschungsgemeinschaft (DFG) through an Emmy Noether Research Group (grant number NE 2441/1-1).
PT acknowledges support from NSF AST-1909933, NSF AST-2008490, and NASA ATP Grant 80NSSC20K0502.
MV acknowledges support through NASA ATP 19-ATP19-0019, 19-ATP19-0020, 19-ATP19-0167, and NSF grants AST-1814053, AST-1814259, AST-1909831, AST-2007355 and AST-2107724.
RW is supported by the Natural Sciences and Engineering Research Council of Canada (NSERC), funding reference CITA 490888-16.
This research was supported in part by the National Science Foundation under Grant No. NSF PHY-1748958.

\section*{Data Availability}

The realistic GC catalogs are available to the public. They can be downloaded here: www.tng-project.org/doppel22 or in the TNG50 public data release upon acceptance of this paper. The catalog of all GC candidates or additional realizations of the realistic GC catalogs will also be made available upon request. 



\bibliographystyle{mnras}
\bibliography{gc_paper_2} 




\appendix

\section{Mass-to-light calibrations}
\label{app:colors}

\begin{figure}
    \centering
    \includegraphics[width = \columnwidth]{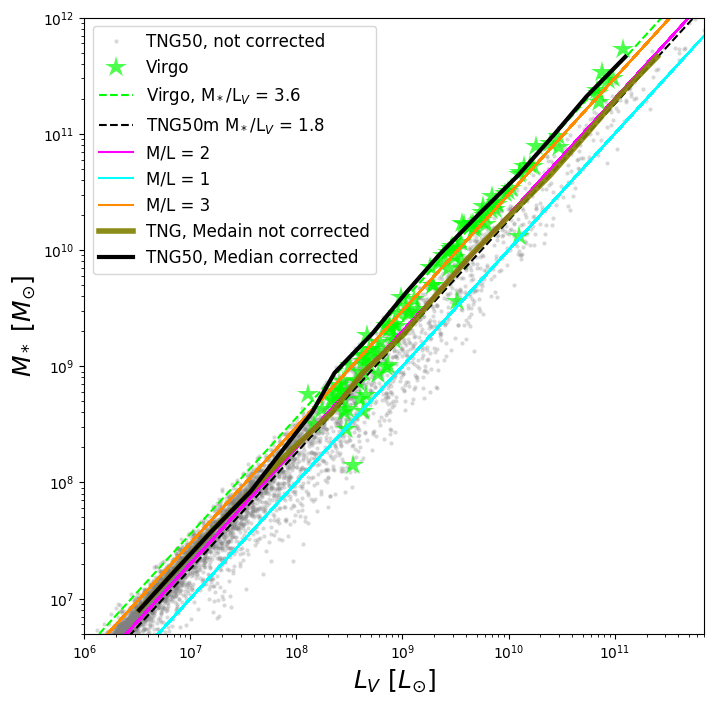}
    \caption{Mass to light ratio for TNG50 (gray points that show individual galaxies, with the gray-green line showing the median bins of stellar mass and the black dotted line showing the best fit) compared to that of Virgo galaxies overplotted as lime green stars (with the best fit shown as the dotted lime green line). At fixed stellar mass, Virgo galaxies tend to be less luminous at higher-masses than simulated objects. This discrepancy at face value in mass-to-light ratio between TNG50 and Virgo for high-mass galaxies causes a discrepancy in both $M_V$ and $S_N$ for those masses. We therefore adopt a ``corrected" mass-to-light ratio (shown in black line) to compute $S_N$ in our results.}
    \label{fig:bcomp}
\end{figure}

Simulated galaxies in TNG50 have stellar masses and corresponding luminosities calculated in several bands, including information on the $V$-band magnitudes necessary, for example, for computing the specific frequency $S_N$ in Fig.~\ref{fig:ngc}. However, the simulated luminosities include only evolution due to stellar population models and might neglect important effects, such as dust attenuation. We therefore compute the $V$-band luminosities of our sample by using a mass-to-light ratio calibration fit to the Virgo cluster data (using stellar masses and V-band absolute magnitudes from \citet{peng2008}). 

This is shown in Fig.~\ref{fig:bcomp}, where gray symbols indicate the results directly from the simulations and green stars are data from Virgo. Thin colored lines indicate constant mass-to-light ratios, as labeled, while the thick black solid line highlights the conversion used in this paper. As expected, the calibrated relation differs from the simulated values mostly at the high-mass end, where dust effects might be playing a more important role. While this correction does not significantly impact any of the results in this paper, considering a mass-to-light ratio equal to $3.6$ for more massive galaxies (instead of $\sim 2$ as suggested by the simulation) improves the agreement with $S_N$ data reported in Sec.~\ref{sec:N_GCs}. 

\begin{figure}
    \centering
    \includegraphics[width = \columnwidth]{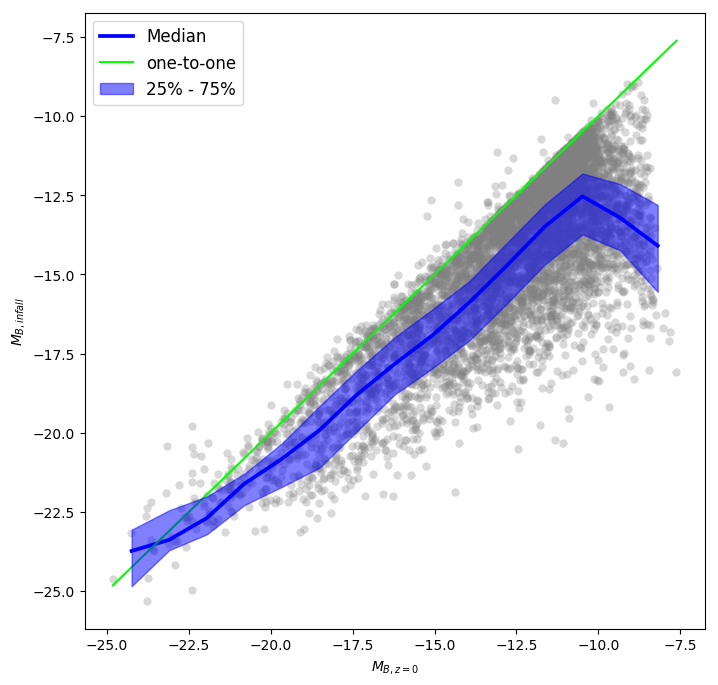}
    \caption{Comparison between galaxy $M_B$ at infall and at $z = 0$. Since observational results ara available at $z=0$ but our tagging occurs at infall, we require a calibration that seeks to take the brighter B-luminosities at infall into account when calculating GC luminosity function.}
    \label{fig:m2lv}
\end{figure}

The evolution in mass-to-light ratio and changes in star formation rate once a galaxy becomes a satellite make necessary an additional calibration in our model. This calibration is related to the dispersion in the luminosity (or mass) function of individual GC masses, $\sigma_z$, described in Sec.~\ref{ssec:gcmass} in our main article and shown in Fig.~\ref{fig:mass_hist} as a function of stellar mass $M_*$. This $z$-band luminosity dispersion $\sigma_z$ is observationally constrained at $z=0$ as a function of the {\it present-day} $B$-band magnitude \citep{jordan2007}. However, the GC tagging and mass assignment in our model is done at infall (and not present day), requiring of an adjustment at the moment to perform the GC tagging to reproduce the desired results at $z=0$. 

For illustration, we show in Fig.~\ref{fig:m2lv} the $B$-band luminosity evolution in all our galaxies from infall to $z=0$. To compensate for this evolution, we first calculate the ``target" relation between $\sigma_z$ and stellar mass $M_*$ (shown in orange in the bottom panel of Fig.~\ref{fig:mass_func}), where $M_*$ is calculated as the median $M_*$ in our simulated galaxies with a given $B$-band luminosity, all at $z=0$. Next, we correct the initial $\sigma_z$ (e.g., at infall time) by calculating the $\sigma_z$ that would correspond to each galaxy assuming their infall stellar mass and then multiply that by a constant factor: $\sigma_{z, \rm inf} = \alpha * \sigma_z(M_*)$, where $\sigma_z(M_*)$ is our target relation at $z=0$ as described before. After experimenting with different values, we find $\alpha=0.75$ a reasonable choice, in particular to reproduce the median $\sigma_z$ at $z=0$ observed in low-mass galaxies, which is the main focus of this work.

\section{Potential Effects of Dynamical Friction}
\label{app:dynamical_friction}

Massive objects such as GCs can experience dynamical friction as they move within the gravitational potential of the smoothly distributed mass in the host galaxy. Our tagging method does not self-consistently follow this effect since we tag them onto the dark matter particles and all components (dark matter, baryons and GCs) have similar particle mass in our simulations. 
By default, our GC catalog ignores dynamical friction effects since the method is tailored to tag only the ``surviving" population of GCs and not the initial one. However, it is important to double-check that after tagging our GCs they would not be substantially affected by dynamical friction and expected to coalesce to the center of the galaxies and be dissolved. 

To gain some intuition, we estimate analytically the typical timescales for dynamical friction in our systems following Eq. 8.17 in \citet{bandt2008}:

$$t_{\rm fric} = \frac{2.7 Gyr}{\rm ln \Lambda} \frac{r_i}{30\; \rm kpc} \left( \frac{\sigma_H}{200\; \rm km/s}\right)^2 \left( \frac{100 \rm km/s}{\sigma_{GC}} \right)^3 ,$$

\noindent
where $\sigma_H$ is the typical velocity dispersion in the host, $\sigma_{\rm GC}$ is the velocity dispersion of the GC, both as proxies for mass, $r_i$ is the initial radius of the GC orbit and $\rm ln \Lambda = 5.8$ is assumed as a typical Coulomb logarithm.  We vary the velocity dispersion of the host assuming $\sigma_H=800, 200, 50, 20$ and $10$ km/s corresponding roughly to the scales of a cluster, a MW like galaxy and dwarfs with $M_* \sim 10^9, 10^8$ and $10^{6.5}$ \msun, respectively. For the GCs, we compare the effects on two scales: a $2 \times 10^5$ \msun\; (typical GC mass) and $1 \times 10^4$ \msun\; (our lower limit and common value in low-mass galaxies). We assume a half-mass radius $r_h = 3$ pc to translate GC mass into velocity dispersion $\sigma_{GC}$. Finally, we consider the radius $r_i$ as the distance of the GC to the center of the host at infall (the moment of the tagging).

\begin{figure*}
    \centering
    \includegraphics[width = \columnwidth]{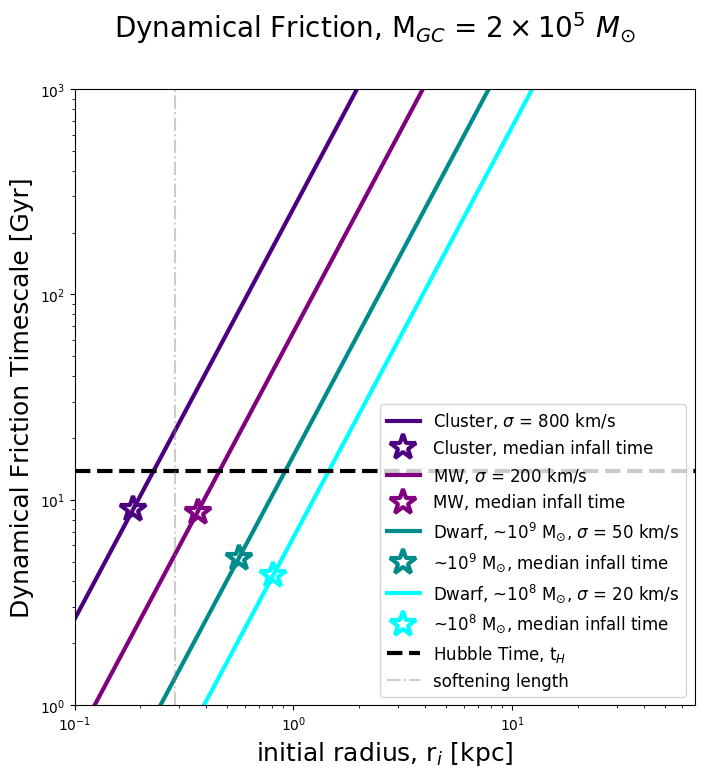}
    \includegraphics[width = \columnwidth]{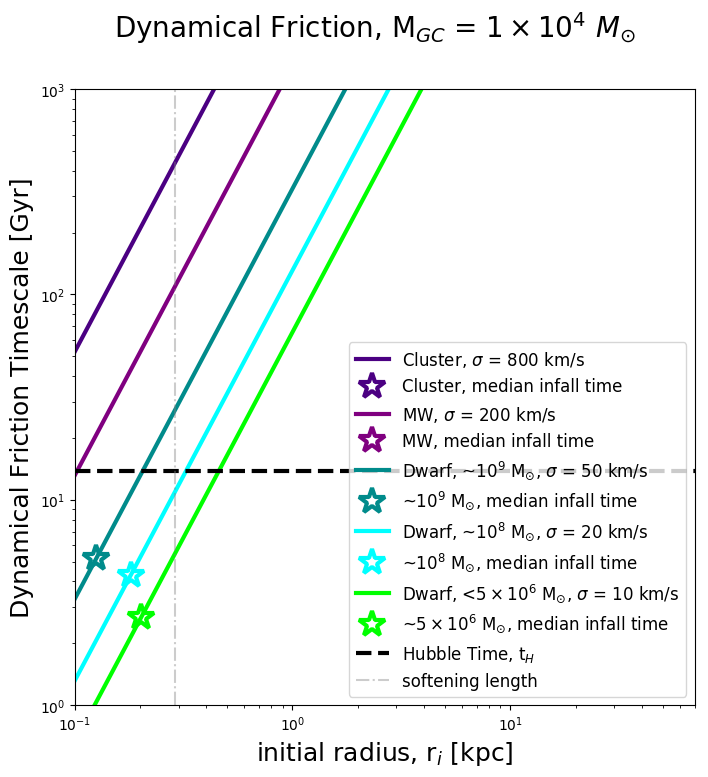}
    \caption{Dynamical friction timescales for different types of GC host systems, assuming the median GC mass of $2\times 10^5$ M$_{\odot}$ (left) and $10^4$ M$_{\odot}$ (right). The typical timescales associated to dynamical friction are longer than the age of the Universe for most initial radii and in particular for high-mass hosts. We also indicate the median infall time for galaxies in each mass range with a starry symbol. For dwarf galaxies, dynamical friction timescales might be lower than a Hubble time only for GCs at very small radii $r<0.5$-$1.0$ kpc, depending on GC mass, but comparable to the time since their infall time, when GCs are tagged. We therefore expect not a significant change in any of the results when including dynamical friction. Notice that our least massive dwarfs do not have GCs as massive as $2\times 10^5 $ M$_{\odot}$ and therefore are not included on the left panel.}
    \label{fig:dynfric_time}
\end{figure*}

We show the results in Fig.~\ref{fig:dynfric_time}, where the dynamical friction timescales are shown as a function of the distance of the GC. For reference, we indicate the age of the Universe with a thick dashed horizontal line, areas where $t_{\rm fric}$ is above the Hubble time $t_H$ indicates that dynamical friction effects are unimportant. As expected, the dynamical friction timescales increase with radius, meaning that only GCs in the very inner regions are potentially affected. Fig.~\ref{fig:dynfric_time} also shows that $t_{\rm fric}$ is shorter for more massive GCs, as expected, but even in this case only GCs within $\sim 1$ kpc have the potential to decay and coalesce due to dynamical friction forces. In the case of a lighter GC, as the one shown on the right panel, the relevant distance where dynamical friction effects might be important shrinks to $\sim 0.5$ kpc. 

Reassuringly, the distances where dynamical friction migth be a factor of concern are quite small compared to the typical GCs radial extension (see Fig.~\ref{fig:rhalf}) and suggest that dynamical friction effects are not important in our sample. Moreover, the time of relevance is not the age of the Universe but the time since infall, when the GC is tagged. Those are highlighted with a starry symbol in Fig.~\ref{fig:dynfric_time} and correspond to the median infall times of galaxies of a given stellar mass in our sample. On average, dynamical friction effects are negligible and if present, may impact only the lowest mass galaxies in the sample.  

\begin{figure}
    \centering
    \includegraphics[width = \columnwidth]{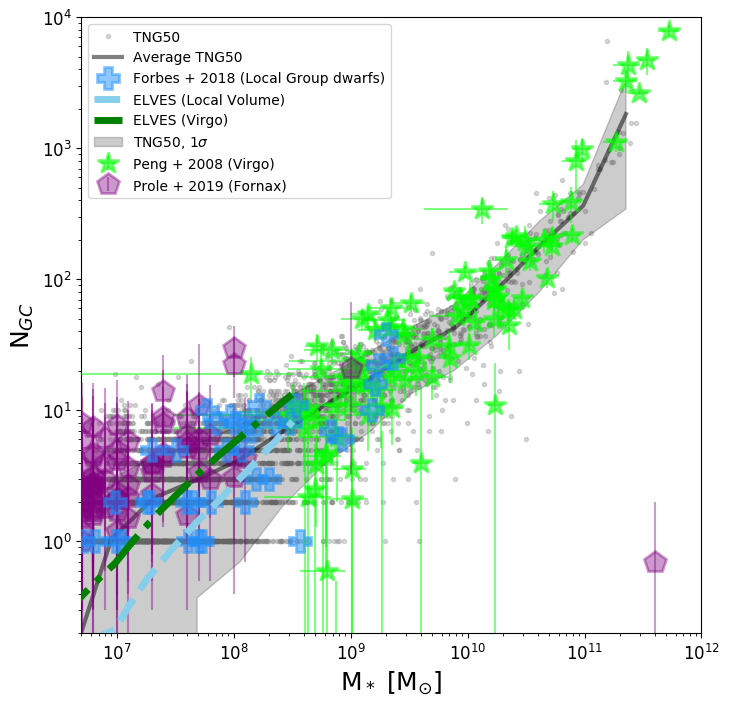}
    \caption{The same as the left panel of Fig \ref{fig:ngc}, but including an estimation of the removal of GCs by dynamical friction. We see very little change in the overall behavior of the GC abundances with stellar mass when including dynamical friction.}
    \label{fig:dynfric_ngc}
\end{figure}

\begin{figure}
    \centering
    \includegraphics[width = \columnwidth]{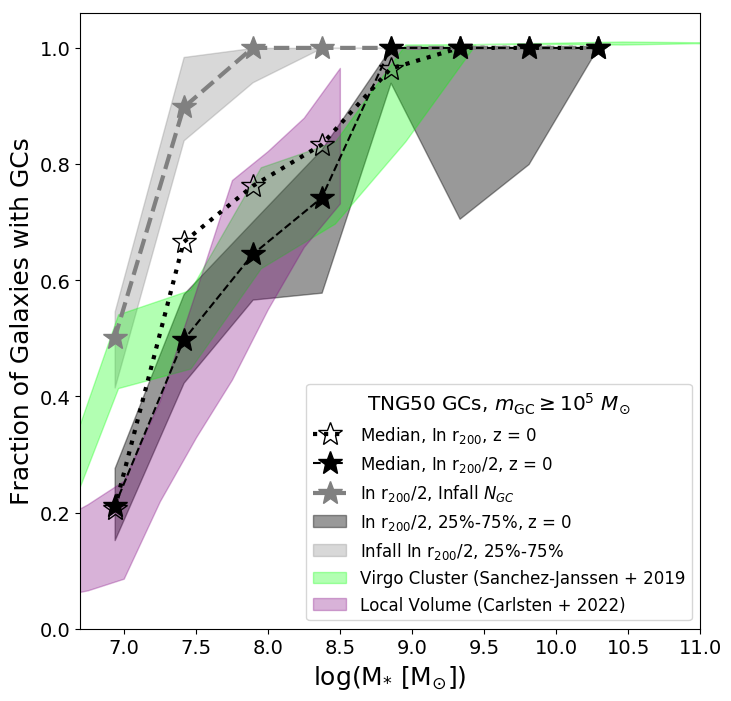}
    \caption{The same as figure \ref{fig:occ_frac}, but including an estimation for removal of GCs by dynamical friction. We see that the dwarf galaxy stellar mass bins that previously sat above observed values fall nicely within the range of observations when including this effect.}
    \label{fig:dynfric_occ}
\end{figure}

Next, using the same equation above, we compute a dynamical friction time individually for each tagged GC and comparing $t_{\rm fric}$ to the particular infall time of that host galaxy we can individually assess whether GCs are expected to decay or not. We flag all GCs where $t_{\rm fric} < (t_{H} - t_{\rm inf})$ as ``merged", and remove them from our sample at $z=0$. Fig.~\ref{fig:dynfric_ngc} and Fig.~\ref{fig:dynfric_occ} show that this would have no significant consequences for our main results, including the number of GCs per galaxy or the occupation fraction, respectively. We therefore conclude that while dynamical friction might impact a few of our GC on an individual basis none of the statistical results presented here changes appreciably. In our released catalog, we provide a dynamical friction flag to allow the user to decide whether to include these objects or not in their calculations.

\section{Environmental Effects}
\label{app:no_environ}

\begin{figure}
    \centering
    \includegraphics[width = \columnwidth]{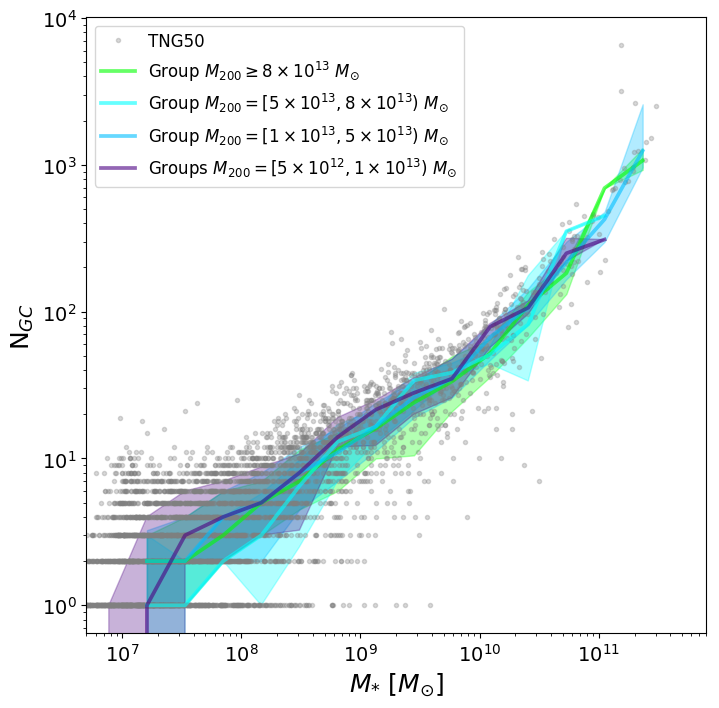}
    \caption{Number of GCs, $N_{\rm GC}$ as a function of host galaxy stellar mass $M_{*}$ binned in host cluster virial mass. The solid lines show the median in bins of host galaxy stellar mass with the shaded region showing the $25\%-75\%$ scatter in each bin. We find no pronounced dependence on GC abundance with host group or cluster environment.}
    \label{fig:ngc_environ_effect}
\end{figure}

Inspired by observations of galaxies in higher density environments showing a higher GC abundance \citep[e.g.][]{peng2008, Carlsten2022}, we have checked if this phenomenon was present in our tagged GCs catalog in TNG50. We split the environments in bins of virial mass, and within those bins, we computed the median and $25\%-75\%$ range of GC abundance in bins of host galaxy stellar mass. Fig. \ref{fig:ngc_environ_effect} shows the result of this test. There is little if any variation in both the median and the scatter between the different virial mass bins. Running the same check on $S_N$ shows the same lack of dependence with the host.

\begin{figure}
    \centering
    \includegraphics[width = \columnwidth]{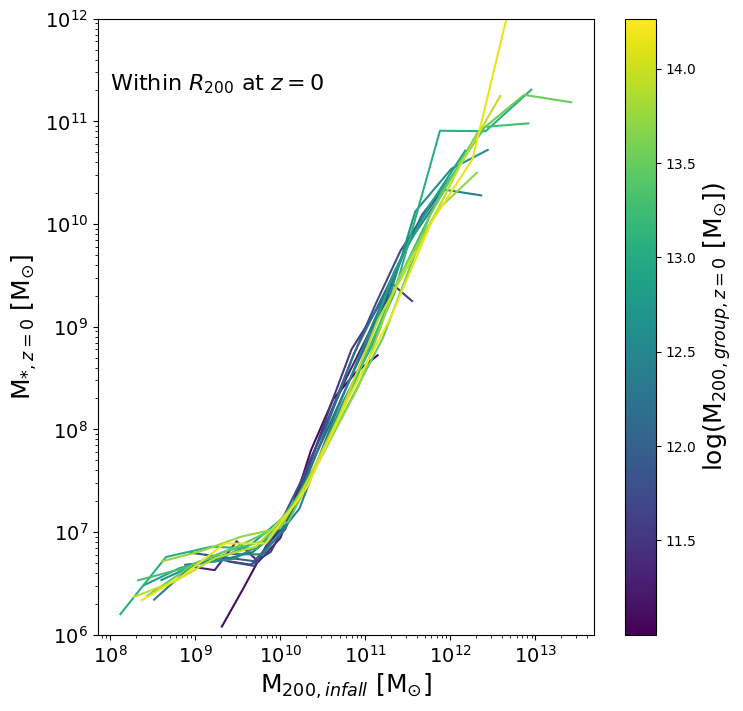}
    \caption{$M_{*}$ at $z = 0$ as a function of infall virial mass $M_{200, \rm infall}$ for galaxies within $R_{200}$ of their $z = 0$ host environment in TNG50. Medians are colored by $z = 0$ host virial mass (color bar on the right). There is a weak but systematic trend for satellites with fixed stellar mass today to have a larger infall virial mass in more massive hosts, in particular for $M_* > 10^7$ \msun. Note that we extend the calculation to host virial masses $M_{200} = 10^{11}$ M$_{\odot}$, which is well below our minimum host halo mass tagged, in order to clearly see the effect. For hosts with $M_{200}>5 \times 10^{12}$ \msun, as studied here, there is not enough difference in satellite infall masses to lead to any environmental trend on GC content.} 
    \label{fig:app_minfall}
\end{figure}

The GC tagging model employed in this work relies on the infall virial mass of a galaxy; thus we checked to see how infall virial mass varies across the tagged environments in Fig.~\ref{fig:app_minfall}. Binning again in host environment virial mass, we calculated the median infall virial mass in bins of {\it present-day} host galaxy stellar mass, $M_{*,\rm z=0}$. We find a weak environmental dependence on the infall halo masses at fixed $z=0$ stellar mass that goes in the direction expected: galaxies of a given stellar mass today had a larger infall virial mass for high density environments \citep[see e.g., ][]{mistani2016}. However, we had had to include low-mass host halos with present-day $M_{200} \geq 10^{11} \ \rm M {\odot}$ (which are not included in our catalog) in order to observe the effect. Limiting the host halo mass to the ones included in this study ($M_{200} > 5 \times 10^{12}$ \msun\;) shrinks the effect appreciably, explaining why our GC catalog shows no significant dependence with environment. Thus, it may be necessary to study a much wider range of host halo masses in order to see the observed environmental dependence on GC abundance.

\bsp	
\label{lastpage}
\end{document}